\documentclass{mn2e} 
\usepackage{graphics}

\def\ni{\noindent}                                       %No indent%
\def\etal{et\thinspace al.\ }                               %et al.%
\def\ojo{\fbox{\bf !$\odot$j$\odot$!}}      %Olho! Needs Correction%

\newcommand{\ov}[1]{\overline{#1}}                        %overline%

\newbox\grsign \setbox\grsign=\hbox{$>$} \newdimen\grdimen \grdimen=\ht\grsign
\newbox\simlessbox \newbox\simgreatbox
\setbox\simgreatbox=\hbox{\raise.5ex\hbox{$>$}\llap
     {\lower.5ex\hbox{$\sim$}}}\ht1=\grdimen\dp1=0pt
\setbox\simlessbox=\hbox{\raise.5ex\hbox{$<$}\llap
     {\lower.5ex\hbox{$\sim$}}}\ht2=\grdimen\dp2=0pt
\def\simgreat{\mathrel{\copy\simgreatbox}}
\def\simless{\mathrel{\copy\simlessbox}}
\title[The evolution of stars and gas in starburst galaxies]
       {The evolution of stars and gas in starburst galaxies}

\author[R. Cid Fernandes, J. R. S. Le\~ao \& R. Rodrigues Lacerda]
	{Roberto Cid Fernandes$^{1}$\thanks{e-mail: cid@astro.ufsc.br},
	Jo\~ao R. S. Le\~ao $^{1}$\thanks{e-mail: joao@fsc.ufsc.br},
	 Reiner Rodrigues Lacerda$^{1}$\thanks{e-mail: reiner@fsc.ufsc.br}\\
  $^1$Depto.\ de F\'{\i}sica - CFM - Universidade Federal de Santa Catarina, 
CP 476, Campus Universit\'ario, Trindade,\\
88040-900, Florian\'opolis, SC, Brazil}

\begin{document}

\maketitle

%AAAAAAAAAAAAAAAAAAAAAAAAAAAAAAAAAAAAAAAAAAAAAAAAAAAAAAAAAAAAAAAAAAA
\begin{abstract} 
In systems undergoing starbursts the evolution of the young stellar
population is expected to drive changes in the emission line
properties. This evolution is usually studied theoretically, with a
combination of evolutionary synthesis models for the spectral energy
distribution of starbursts and photoionization calculations. In this
paper we present a more empirical approach to this issue.  We apply
empirical population synthesis techniques to samples of Starburst and
HII galaxies in order to measure their evolutionary state and
correlate the results with their emission line properties.  A couple
of useful tools are introduced which greatly facilitate the
interpretation of the synthesis: (1) an evolutionary diagram, whose
axis are the strengths of the young, intermediate age and old
components of the stellar population mix, and (2) the mean age of
stars associated with the starburst, $\ov{t}_{SB}$. These tools are
tested with grids of theoretical galaxy spectra and found to work very
well even when only a small number of observed properties (absorption
line equivalent widths and continuum colors) is used in the synthesis.

Starburst nuclei and HII galaxies are found to lie on a well defined
sequence in the evolutionary diagram. Using the empirically defined
mean starburst age in conjunction with emission line data we have
verified that the equivalent widths of H$\beta$ and [OIII] decrease
for increasing $\ov{t}_{SB}$. The same evolutionary trend was
identified for line ratios indicative of the gas excitation, although
no clear trend was identified for metal rich systems.  All these
results are in excellent agreement with long known, but little tested,
theoretical expectations.
\end{abstract}

\begin{keywords}
galaxies: starburst - galaxies: evolution - galaxies: stellar content
- ISM: HII Regions
\end{keywords}
%AAAAAAAAAAAAAAAAAAAAAAAAAAAAAAAAAAAAAAAAAAAAAAAAAAAAAAAAAAAAAAAAAAA

\section{Introduction}

\label{sec:Introduction}

Starburst systems in the local universe, from giant HII regions to
Starburst nuclei, are important laboratories to study the evolution of
massive stars and physical processes thought to be associated with the
very early stages of galaxy formation. These motivations, coupled to
the advances in modeling and observational capabilities, have led to
a burst of activity in this field during the past decade.

By far the most common approach to infer the physical properties of
starbursts is to compare their spectral energy distribution (SED) with
models based on evolutionary synthesis (eg.\ Mas-Hesse \& Kunth 1991;
Olofsson 1995; Leitherer \etal 1999). This technique performs {\it ab
initio} calculations of the spectral evolution of a stellar population
on the basis of evolutionary tracks for stars covering a wide range of
masses, stellar spectral libraries plus prescriptions for the initial
mass function, star formation rate and chemical evolution. The
comparison between observations and models may focus on particular
spectral features, such as stellar wind lines in the UV (Robert,
Leitherer \& Heckman 1993), WR features (Cervi\~no \& Mas-Hesse 1994;
Schaerer \& Vacca 1998), Balmer absorption lines (Gonz\'alez Delgado,
Leitherer \& Heckman 1999), lines from red super giants
(Garc\'{\i}a-Vargas \etal 1997; Mayya 1997), or on a combination of
lines and the multiwavelength continuum (Mas-Hesse \& Kunth 1999;
Lan\c{c}on \etal 2001). A common difficulty faced in such studies is
the contamination of the spectrum by an underlying old stellar
population, which can be significant in the optical--near IR
range. This contamination is sometimes removed by adopting a template
spectrum for the old population (Lan\c{c}on \etal 2001), or else its
effects are evaluated from the excess flux between models and data
(Mas-Hesse \& Kunth 1999).  Other uncertainties include those
associated with differences between different sets of evolutionary
tracks, incomplete or imperfect spectral libraries; stochastic effects
and numerical issues (Cervi\~no \etal 2001); and possible effects of
binary stars (Mas-Hesse \& Cervi\~no 1999). We refer the reader to the
review by Schaerer (2001) for a detailed discussion.

The massive, hot stars in young starbursts photoionize the surrounding
gas, producing an emission line spectrum which can also be used to
constrain the properties of the starburst, like its age, metallicity
and star-formation rate. In fact, emission line diagnostics of
starbursts play a central role in this field for the simple practical
reason that emission lines are much easier to measure than stellar
features.  A great number of studies have been devoted to developing
such diagnostics through a combination of theoretical SEDs from
evolutionary synthesis and phototoionization calculations for the
corresponding nebular conditions (Garc\'{\i}a-Vargas, Bressan \&
D\'{\i}az 1995; Stasi\'nska \& Leitherer 1996; Charlot \& Longhetti
2001; Moy, Rocca-Volmerange \& Fioc 2001; Stasi\'nska, Schaerer \&
Leitherer 2001; Schaerer 2000).

Among the many diagnostic tools developed, the equivalent width of
H$\beta$ ($W_{H\beta}$) stands out as a powerful {\it age indicator}
of starbursts. The prediction, known for more than 20 years (Dottori
1981), is that $W_{H\beta}$ decreases as a burst evolves.  This
behavior has been so extensively confirmed by more elaborate
calculations that $W_{H\beta}$ is often used as a substitute for an
age axis in studies which investigate the evolution of starbursts
(e.g., Stasi\'nska \etal 2001).  Other general predictions of
evolutionary synthesis + photoionization models are that the gas
excitation and that the equivalent width of [OIII]$\lambda$5007 should
decrease as a starburst evolves (Coppeti, Dottori \& Pastoriza 1986;
Stasi\'nska \& Leitherer 1996) although such diagnostics are more
critically affected by metallicity effects.

These predictions, now routinely applied to infer physical properties
of starbursts, are hard to be tested directly, since that requires
evaluating the age of a starburst without resorting to emission line
diagnostics. In this paper we take a step back in time and investigate
the empirical validity of these long known predictions by means of a
simple Empirical Population Synthesis (EPS) analysis of Starburst and
HII galaxies.  This analysis allows a quantitative assessment of the
evolutionary state of a stellar population based only on observed
stellar features. EPS techniques have their own limitations (eg, Cid
Fernandes \etal 2001a), but these are of a different nature than the
uncertainties involved in evolutionary synthesis, and thus serve as an
independent test of predictions of evolutionary synthesis models.

Our main goals are to:	

\begin{itemize}

\item[(i)] Study the stellar population properties of a large and
varied sample of star-forming galaxies by means of an EPS analysis.

\item[(ii)] Develop and test EPS-based tools to assess the
evolutionary state of star-forming galaxies in a quantitative and
easy-to-interpret manner.

\item[(iii)] Perform a model-independent investigation of the relation
between gaseous properties and the evolutionary state of starbursts.

\end{itemize}

%This paper is organized as follows. 
In section~\ref{sec:Data} we present the data sets used in this
study. Section \ref{sec:Synthesis} deals with points {\it (i)} and
{\it (ii)} above. The EPS method is presented and its results for
star-forming galaxies are discussed by means of simple empirical tools
designed to aid the interpretation of the synthesis.  We also present
a comparative study of EPS and evolutionary synthesis methods, which
serves to test and calibrate our tools to measure evolution.  In
section \ref{sec:Emission_Line_Properties} we address point {\it
(iii)} by studying the evolution of emission line properties of
Starburst and HII galaxies using an EPS-based measure of the burst
age.  Section \ref{sec:conclusions} summarizes our main results.

\section{Data}

\label{sec:Data}

This investigation requires optical spectra where both gaseous
(emission lines) and stellar properties (continuum and absorption
lines) can be discerned. Two data sets meeting this requirement were
used in this work.

The first set, which we denote ``Sample I'', comes from the studies of
Storchi-Bergmann, Kinney \& Challis (1995) and McQuade, Kinney \&
Calzetti (1995), extracted from the atlas of Kinney \etal (1993). This
sample covers both large, luminous galaxies with star-forming nuclei
(``Starburst nuclei'') and smaller, weaker systems such as HII
galaxies and blue compact dwarves.  Active galaxies were discarded,
with the exception of NGC 6221, which is dominated by Starburst
activity except in X-rays (Levenson \etal 2001). We further limit our
analysis to those objects with metallicity estimates by
Storchi-Bergmann, Calzetti \& Kinney (1994), which leaves a total of
41 galaxies, 18 of which are classified as Starburst nuclei. The
spectra were collected through a large $10^{\prime\prime} \times
20^{\prime\prime}$ aperture, which corresponds to a circular aperture
of 1.3 $h_{75}^{-1}$ kpc in radius at the median distance of the
galaxies.

Emission line fluxes and equivalent widths ($W$) for this sample were
re-measured from the publicly available original spectra and found to
be in good agreement with those obtained by Storchi-Bergmann \etal
(1995) and McQuade \etal (1995).  All spectra were corrected for
Galactic extinction using the reddening law of Cardelli, Clayton \&
Mathis (1989, with $R_V = 3.1$) and the $E(B-V)$ values from Schlegel,
Finkbeiner \& Davis (1998) as listed in NED\footnote{The NASA/IPAC
Extragalactic Database (NED) is operated by the Jet Propulsion
Laboratory, California Institute of Technology, under contract with
the National Aeronautics and Space Administration.}. Corrections for
internal extinction were applied based on the H$\alpha$/H$\beta$
ratio, whose intrinsic value was taken to be 2.86, and allowing for
underlying absorption components (see
Section~\ref{sec:W_Hb_corrections}). For the stellar population
analysis we have measured the $W$'s of the CaII~K~$\lambda$3933,
CN~$\lambda$4200 and G~band~$\lambda$4301 with respect to a pseudo
continuum defined at selected pivot points, located at $\lambda =
3660$, 3780, 4020 and 4510 \AA, following the methodology outlined in
Cid Fernandes, Storchi-Bergmann \& Schmitt (1998).  Our values agree
very well with those of Storchi-Bergmann \etal (1995), but there were
significant discrepancies between our measurements and those published
by McQuade \etal (1995).

``Sample II'' comes from the Spectrophotometric Atlas of HII galaxies
of Terlevich \etal (1991), as analysed by Raimann \etal
(2000a,b). Most of the individual spectra in this atlas do not have
enough signal to measure stellar features, which prompted Raimann
\etal to average them in order to increase the stellar signal. Out of
185 galaxies, they have defined 19 {\it groups} of similar
characteristics.  Each group is then treated as if it corresponded to
an individual galaxy.  Three of these groups are composed of Seyfert 2
galaxies. These were kept in our analysis only to illustrate their
systematically different properties with respect to the remaining
groups, 10 of which are composed of HII galaxies, 2 are Starburst
nuclei and 4 are classified as intermediate HII/Staburst systems. The
typical aperture covered by these spectra correspond to an equivalent
radius of $0.8 h_{75}^{-1}$ kpc.

Raimann \etal (2000a) have measured absorption line $W$'s and
continuum fluxes for these groups following the same methodology as
above.  Emission line properties were analysed by Raimann \etal
(2000b). Line fluxes were initially measured after subtraction of a
stellar population model which included internal reddening. The
resulting H$\alpha$/H$\beta$ ratio was used to further correct for
residual extinction towards the line emitting regions.  We have adopted
both stellar and nebular properties as published by these authors
without further corrections.

In summary, of all stellar and nebular properties compiled for these 2
samples, the following will be used in the analysis below: (1) the
$W$'s of the CaII K, CN and G-band absorption features; (2) continuum
fluxes at 3600, 4020 and 4510 \AA; (3) emission line fluxes and $W$'s
of strong optical lines: [OII]$\lambda$3727, H$\beta$,
[OIII]$\lambda$5007, H$\alpha$ and [NII]$\lambda$6584; (4) nebular
abundances, as listed in Storchi-Bergmann \etal (1994) and Raimann
\etal (2000b); (5) the ``activity class'', as reported in the afore
mentioned papers. This last item is used to distinguish small systems
like HII and blue compact galaxies from Starburst nuclei, which live
on larger and more luminous galaxies, usually late type spirals. The
latter kind of galaxies present a more complex mixture of stellar
populations and are richer in heavy elements than HII galaxies, as
will become clear in the analysis that follows.

\section{Empirical Population Synthesis analysis}

\label{sec:Synthesis}

\subsection{The method: input and output quantities}

\label{sec:Synthesis_Method}

In order to provide a quantitative description of the stellar
populations for galaxies in Samples I and II, we have used their
absorption line $W$'s and continuum colors as input to the EPS
algorithm developed by Cid Fernandes \etal (2001a). The code
decomposes a spectrum onto a base of 12 simple stellar populations of
different ages and metallicities ($Z$). This base was defined by
Schmidt \etal (1991) out of a large sample of star clusters originally
observed by Bica \& Alloin (1986a,b).  The main output of the code is
the {\it population vector} ${\bf x}$, whose 12 components carry the
fractional contributions of each base element to the observed flux at
a given normalization wavelength $\lambda_0$.  This vector corresponds
to the {\it mean} solution found from a $10^8$ steps likelihood-guided
Metropolis walk through the parameter space. Since colors are also
modeled, extinction enters as an extra parameter, but this will not be
directly used in our analysis. Some EPS studies (e.g., Bica 1988)
impose that solutions follow well behaved paths on the age-$Z$ plane
spanned by the base, in order to force consistency with simple
scenarios for chemical evolution. Here we follow Schmidt \etal (1991)
in not imposing such {\it a priori} constraints in order to allow for
more general scenarios, such as systems undergoing mergers.

There are no major conceptual differences between this EPS method and
that originally developed by Bica (1988) or its variants, which have
been applied to many stellar population studies (e.g., Bica, Alloin \&
Schmidt 1990; De Mello \etal 1995; Kong \& Cheng 1999; Schmitt,
Storchi-Bergmann \& Cid Fernandes 1999; Raimann \etal 2000a). However,
in this work we will explore novel ways of expressing the results of
the synthesis, which use the population vector to construct
easy-to-interpret diagrams and indices.

The results presented below were all obtained feeding the EPS code
with just 5 observables: The $W$'s of CaII K, CN and the G-band, plus
the $F_{3660}/F_{4020}$ and $F_{4510}/F_{4020}$ continuum colors. The
errors on these quantities were fixed at 0.5 \AA\ for $W_K$ and $W_G$,
1 \AA\ for $W_{CN}$, and 0.05 for the colors. As discussed by Cid
Fernandes \etal (2001a), the combination of observational errors,
little input information and quasi-linear dependences within the base
hinders accurate estimates of all 12 components of ${\bf x}$, but
reliable results are obtained grouping ${\bf x}$ components of {\it
same age}.  We have therefore employed age-grouping schemes in our
analysis.

The base spans 5 logarithmicaly spaced age bins: $10^6$, $10^7$,
$10^8$, $10^9$ and $10^{10}$ yr. Components with these ages are
combined onto a reduced 5-D population vector, whose components are
denoted by $x_6$, $x_7$, $x_8$, $x_9$ and $x_{10}$ respectively.  We
will also work with an even further reduced description of stellar
populations, in which the $10^9$ and $10^{10}$ yr old components are
re-grouped onto $x_O \equiv x_9 + x_{10}$, and the young $10^6$ and
$10^7$ yr components are binned onto $x_Y \equiv x_6 + x_7$.  Renaming
the ``intermediate age'' $10^8$ yr component to $x_I \equiv x_8$, we
obtain a compact, 3-D version of the population vector:
$(x_Y,x_I,x_O)$.

Normalization requires that $x_Y + x_I + x_O = 1$, while the
positivity constraint implies that all components are $\ge 0$.
Therefore, an EPS solution is confined to a triangular cut of a plane
in the $(x_Y,x_I,x_O)$ space, which facilitates the visualization of
results (see Cid Fernandes \etal 2001b for an application of this
scheme to trace the evolution of circumnuclear starbursts in active
galaxies). We note that the description of a galaxy spectrum in terms
of ${\bf x}$-components is analogous to a Principal Component Analysis
(e.g., Sodr\'e \& Stasi\'nska 1999), with the difference that, by
construction, each component has a known physical meaning.

We concentrate our analysis of the EPS results on evolutionary
effects. Furthermore, we focus on the evolution of recent stellar
generations ($\le 10^8$ yr), associated with the star-forming activity
in Starburst and HII galaxies. Metallicity effects are discussed using
the nebular oxygen abundance, which essentially reflects the
metallicity of the most recent stellar generation (Storchi-Bergmann
\etal 1994).

\subsection{EPS results and the evolutionary diagram}

\label{sec:EPS_Results_for_Samples_I_and_II}

%***TAB***TAB***TAB***TAB***TAB***TAB***TAB***TAB***TAB***TAB***TAB
\begin{table*}
\begin{centering}
\begin{tabular}{lrrrrrrrr}
\multicolumn{9}{c}{EPS Results for Samples I and II}\\ \hline
Galaxy          &
$x_6$           &
$x_7$           &   
$x_8 = x_I$     & 
$x_9$           & 
$x_{10}$        & 
$x_Y$           & 
$x_O$           & 
$\log \ov{t}_{SB}$ [yr] \\ \hline
ESO 296-11        & 22 $\pm$  7 & 15 $\pm$  8 & 29 $\pm$  7 & 10 $\pm$  5 & 23 $\pm$  8 & 38 $\pm$  6 & 33 $\pm$  8 &     7.1 $\pm$  0.9 \\ 
ESO 572-34        & 28 $\pm$  9 & 29 $\pm$ 11 & 16 $\pm$  6 & 12 $\pm$  5 & 15 $\pm$  5 & 57 $\pm$  5 & 27 $\pm$  4 &     6.8 $\pm$  0.5 \\ 
1050+04           & 12 $\pm$  6 & 12 $\pm$  7 & 47 $\pm$  8 & 11 $\pm$  5 & 17 $\pm$  6 & 24 $\pm$  6 & 28 $\pm$  6 &     7.5 $\pm$  0.7 \\ 
Haro 15           & 16 $\pm$  7 & 21 $\pm$  9 & 49 $\pm$  7 &  6 $\pm$  3 &  8 $\pm$  5 & 37 $\pm$  6 & 14 $\pm$  5 &     7.4 $\pm$  0.5 \\ 
IC 1586           & 13 $\pm$  7 & 22 $\pm$  9 & 27 $\pm$  7 & 21 $\pm$  6 & 18 $\pm$  7 & 35 $\pm$  6 & 38 $\pm$  6 &     7.2 $\pm$  0.8 \\ 
IC 214            & 11 $\pm$  6 & 15 $\pm$  7 & 37 $\pm$  7 & 16 $\pm$  6 & 22 $\pm$  5 & 26 $\pm$  5 & 38 $\pm$  5 &     7.4 $\pm$  0.6 \\ 
Mrk 66            & 24 $\pm$  6 & 10 $\pm$  6 & 41 $\pm$  7 &  7 $\pm$  4 & 17 $\pm$  4 & 34 $\pm$  5 & 24 $\pm$  5 &     7.2 $\pm$  0.5 \\ 
Mrk 309           & 16 $\pm$  8 & 32 $\pm$ 10 & 25 $\pm$  7 & 16 $\pm$  5 & 12 $\pm$  5 & 48 $\pm$  6 & 28 $\pm$  5 &     7.1 $\pm$  0.5 \\ 
Mrk 357           & 39 $\pm$ 10 & 31 $\pm$ 12 &  5 $\pm$  3 & 13 $\pm$  5 & 11 $\pm$  4 & 70 $\pm$  4 & 25 $\pm$  3 &     6.5 $\pm$  0.4 \\ 
Mrk 499           & 16 $\pm$  7 & 16 $\pm$  8 & 50 $\pm$  7 &  8 $\pm$  4 & 10 $\pm$  5 & 32 $\pm$  6 & 18 $\pm$  5 &     7.4 $\pm$  0.5 \\ 
Mrk 542           & 11 $\pm$  6 & 16 $\pm$  7 & 49 $\pm$  8 & 10 $\pm$  5 & 15 $\pm$  6 & 27 $\pm$  6 & 24 $\pm$  6 &     7.5 $\pm$  0.7 \\ 
NGC 1140          & 20 $\pm$  8 & 26 $\pm$ 10 & 35 $\pm$  7 &  7 $\pm$  4 & 12 $\pm$  6 & 46 $\pm$  6 & 19 $\pm$  6 &     7.2 $\pm$  0.6 \\ 
NGC 1313          & 16 $\pm$  7 & 19 $\pm$  9 & 50 $\pm$  7 &  5 $\pm$  3 &  9 $\pm$  5 & 35 $\pm$  6 & 14 $\pm$  5 &     7.4 $\pm$  0.5 \\ 
NGC 1510          &  8 $\pm$  5 & 13 $\pm$  6 & 57 $\pm$  7 & 10 $\pm$  5 & 11 $\pm$  5 & 21 $\pm$  5 & 22 $\pm$  5 &     7.6 $\pm$  0.6 \\ 
NGC 1569          & 21 $\pm$  9 & 44 $\pm$ 11 &  3 $\pm$  2 & 25 $\pm$  4 &  8 $\pm$  4 & 65 $\pm$  4 & 32 $\pm$  3 &     6.7 $\pm$  0.4 \\ 
NGC 1614          & 16 $\pm$  7 & 20 $\pm$  9 & 39 $\pm$  8 &  8 $\pm$  4 & 17 $\pm$  8 & 36 $\pm$  6 & 25 $\pm$  7 &     7.3 $\pm$  0.7 \\ 
NGC 1705          & 20 $\pm$  9 & 34 $\pm$ 11 & 30 $\pm$  7 &  7 $\pm$  3 &  9 $\pm$  5 & 54 $\pm$  6 & 16 $\pm$  5 &     7.1 $\pm$  0.5 \\ 
NGC 1800          & 10 $\pm$  6 & 14 $\pm$  7 & 55 $\pm$  8 &  9 $\pm$  4 & 13 $\pm$  6 & 24 $\pm$  6 & 22 $\pm$  6 &     7.6 $\pm$  0.6 \\ 
NGC 3049          &  8 $\pm$  6 & 31 $\pm$  8 & 16 $\pm$  6 & 26 $\pm$  7 & 20 $\pm$  8 & 39 $\pm$  6 & 46 $\pm$  6 &     7.1 $\pm$  0.9 \\ 
NGC 3125          & 21 $\pm$  9 & 40 $\pm$ 11 & 17 $\pm$  6 & 10 $\pm$  4 & 11 $\pm$  5 & 61 $\pm$  6 & 22 $\pm$  5 &     7.0 $\pm$  0.5 \\ 
NGC 3256          & 19 $\pm$  8 & 29 $\pm$ 11 & 36 $\pm$  7 &  6 $\pm$  3 & 10 $\pm$  5 & 48 $\pm$  6 & 16 $\pm$  5 &     7.2 $\pm$  0.5 \\ 
NGC 4194          & 10 $\pm$  6 & 16 $\pm$  7 & 39 $\pm$  7 & 19 $\pm$  6 & 16 $\pm$  6 & 26 $\pm$  5 & 35 $\pm$  6 &     7.4 $\pm$  0.7 \\ 
NGC 4385          &  7 $\pm$  5 & 26 $\pm$  8 & 23 $\pm$  7 & 24 $\pm$  7 & 20 $\pm$  8 & 33 $\pm$  6 & 44 $\pm$  7 &     7.3 $\pm$  0.9 \\ 
NGC 5236          & 10 $\pm$  7 & 30 $\pm$  9 & 35 $\pm$  7 & 11 $\pm$  5 & 13 $\pm$  6 & 40 $\pm$  6 & 25 $\pm$  6 &     7.3 $\pm$  0.6 \\ 
NGC 5253          & 21 $\pm$ 10 & 46 $\pm$ 12 & 16 $\pm$  6 &  8 $\pm$  4 & 10 $\pm$  5 & 67 $\pm$  6 & 18 $\pm$  5 &     6.9 $\pm$  0.5 \\ 
NGC 5860          & 12 $\pm$  6 & 18 $\pm$  8 & 22 $\pm$  7 & 18 $\pm$  8 & 30 $\pm$  9 & 30 $\pm$  6 & 49 $\pm$  8 &     7.2 $\pm$  1.1 \\ 
NGC 5996          &  8 $\pm$  5 & 20 $\pm$  7 &  7 $\pm$  4 & 28 $\pm$  9 & 38 $\pm$  9 & 28 $\pm$  5 & 66 $\pm$  6 &     7.0 $\pm$  1.2 \\ 
NGC 6052          & 10 $\pm$  6 & 27 $\pm$  9 & 32 $\pm$  7 & 15 $\pm$  6 & 16 $\pm$  7 & 37 $\pm$  6 & 31 $\pm$  6 &     7.3 $\pm$  0.7 \\ 
NGC 6090          & 10 $\pm$  6 & 20 $\pm$  8 & 32 $\pm$  7 & 29 $\pm$  5 &  9 $\pm$  4 & 30 $\pm$  5 & 38 $\pm$  4 &     7.4 $\pm$  0.5 \\ 
NGC 6217          &  9 $\pm$  6 & 26 $\pm$  8 & 20 $\pm$  7 & 19 $\pm$  7 & 26 $\pm$  9 & 35 $\pm$  6 & 45 $\pm$  7 &     7.2 $\pm$  1.0 \\ 
NGC 6221          &  6 $\pm$  5 & 19 $\pm$  7 & 26 $\pm$  8 & 24 $\pm$  8 & 24 $\pm$  9 & 25 $\pm$  6 & 48 $\pm$  7 &     7.4 $\pm$  1.1 \\ 
NGC 7250          & 20 $\pm$  9 & 34 $\pm$ 11 & 25 $\pm$  7 & 12 $\pm$  4 &  9 $\pm$  4 & 54 $\pm$  6 & 21 $\pm$  4 &     7.1 $\pm$  0.4 \\ 
NGC 7496          & 18 $\pm$  7 & 17 $\pm$  8 & 38 $\pm$  8 & 10 $\pm$  5 & 17 $\pm$  7 & 35 $\pm$  6 & 27 $\pm$  7 &     7.3 $\pm$  0.7 \\ 
NGC 7552          &  7 $\pm$  5 & 15 $\pm$  7 & 48 $\pm$  8 & 12 $\pm$  5 & 18 $\pm$  8 & 22 $\pm$  6 & 30 $\pm$  8 &     7.6 $\pm$  0.8 \\ 
NGC 7673          & 11 $\pm$  6 & 19 $\pm$  8 & 52 $\pm$  8 &  6 $\pm$  3 & 12 $\pm$  7 & 30 $\pm$  6 & 19 $\pm$  7 &     7.5 $\pm$  0.6 \\ 
NGC 7714          & 17 $\pm$  8 & 27 $\pm$ 10 & 33 $\pm$  7 & 11 $\pm$  5 & 12 $\pm$  6 & 44 $\pm$  6 & 23 $\pm$  6 &     7.2 $\pm$  0.6 \\ 
NGC 7793          &  5 $\pm$  4 & 12 $\pm$  6 & 28 $\pm$  8 & 23 $\pm$  8 & 32 $\pm$ 10 & 17 $\pm$  6 & 55 $\pm$  8 &     7.5 $\pm$  1.4 \\ 
1941-543          & 23 $\pm$  8 & 20 $\pm$  9 & 40 $\pm$  7 &  8 $\pm$  4 &  9 $\pm$  4 & 43 $\pm$  5 & 17 $\pm$  4 &     7.2 $\pm$  0.4 \\ 
Tol 1924-416      & 10 $\pm$  7 & 45 $\pm$  9 &  9 $\pm$  5 & 19 $\pm$  6 & 16 $\pm$  6 & 56 $\pm$  6 & 35 $\pm$  5 &     7.0 $\pm$  0.6 \\ 
UGC 9560          & 31 $\pm$  9 & 24 $\pm$ 10 & 13 $\pm$  6 & 17 $\pm$  6 & 15 $\pm$  4 & 55 $\pm$  5 & 32 $\pm$  4 &     6.7 $\pm$  0.5 \\ 
UGCA 410          & 46 $\pm$  6 &  9 $\pm$  6 &  4 $\pm$  3 &  6 $\pm$  3 & 35 $\pm$  3 & 55 $\pm$  3 & 41 $\pm$  2 &     6.3 $\pm$  0.3 \\ 
\hline
G\_Cam1148-2020   & 91 $\pm$  4 &  6 $\pm$  4 &  1 $\pm$  1 &  1 $\pm$  1 &  2 $\pm$  1 & 97 $\pm$  1 &  3 $\pm$  1 &     6.1 $\pm$  0.1 \\ 
G\_UM461          & 84 $\pm$  7 & 11 $\pm$  7 &  1 $\pm$  1 &  2 $\pm$  1 &  3 $\pm$  1 & 95 $\pm$  2 &  4 $\pm$  2 &     6.1 $\pm$  0.1 \\ 
G\_Tol1924-416    & 59 $\pm$ 11 & 29 $\pm$ 12 &  5 $\pm$  3 &  3 $\pm$  2 &  4 $\pm$  2 & 88 $\pm$  4 &  7 $\pm$  3 &     6.4 $\pm$  0.3 \\ 
G\_NGC1487        & 56 $\pm$ 10 & 29 $\pm$ 12 &  8 $\pm$  4 &  2 $\pm$  2 &  4 $\pm$  3 & 85 $\pm$  5 &  7 $\pm$  3 &     6.5 $\pm$  0.3 \\ 
G\_Tol1004-296    & 47 $\pm$ 11 & 40 $\pm$ 12 &  4 $\pm$  3 &  4 $\pm$  2 &  5 $\pm$  3 & 87 $\pm$  4 &  9 $\pm$  3 &     6.5 $\pm$  0.3 \\ 
G\_UM488          & 42 $\pm$ 11 & 42 $\pm$ 12 &  6 $\pm$  4 &  4 $\pm$  2 &  6 $\pm$  3 & 84 $\pm$  5 &  9 $\pm$  3 &     6.6 $\pm$  0.3 \\ 
G\_Tol0440-381    & 45 $\pm$  9 & 23 $\pm$ 11 & 22 $\pm$  6 &  4 $\pm$  2 &  7 $\pm$  4 & 67 $\pm$  6 & 11 $\pm$  4 &     6.7 $\pm$  0.4 \\ 
G\_UM504          & 35 $\pm$  9 & 21 $\pm$ 10 & 25 $\pm$  7 &  6 $\pm$  3 & 13 $\pm$  7 & 56 $\pm$  6 & 19 $\pm$  6 &     6.9 $\pm$  0.6 \\ 
G\_UM71           & 22 $\pm$  8 & 19 $\pm$  9 & 38 $\pm$  7 &  7 $\pm$  4 & 14 $\pm$  7 & 41 $\pm$  6 & 21 $\pm$  7 &     7.2 $\pm$  0.7 \\ 
G\_NGC1510        & 16 $\pm$  7 & 17 $\pm$  9 & 51 $\pm$  8 &  4 $\pm$  3 & 12 $\pm$  7 & 33 $\pm$  6 & 16 $\pm$  7 &     7.4 $\pm$  0.6 \\ 
G\_Cam0949-2126  & 20 $\pm$  8 & 21 $\pm$  9 & 30 $\pm$  7 & 13 $\pm$  5 & 17 $\pm$  6 & 41 $\pm$  6 & 29 $\pm$  6  &     7.1 $\pm$  0.7 \\ 
G\_Mrk711         & 24 $\pm$ 10 & 38 $\pm$ 12 & 15 $\pm$  6 &  8 $\pm$  4 & 15 $\pm$  7 & 62 $\pm$  7 & 23 $\pm$  6 &     6.9 $\pm$  0.7 \\ 
G\_UM140          & 25 $\pm$  8 & 16 $\pm$  9 & 44 $\pm$  7 &  4 $\pm$  3 & 11 $\pm$  6 & 41 $\pm$  6 & 15 $\pm$  6 &     7.2 $\pm$  0.6 \\ 
G\_NGC3089        & 20 $\pm$  7 & 12 $\pm$  7 & 44 $\pm$  8 &  5 $\pm$  3 & 19 $\pm$  9 & 32 $\pm$  6 & 24 $\pm$  8 &     7.3 $\pm$  0.8 \\ 
G\_Mrk710         & 58 $\pm$ 11 & 30 $\pm$ 12 &  4 $\pm$  3 &  3 $\pm$  2 &  5 $\pm$  3 & 87 $\pm$  4 &  8 $\pm$  3 &     6.4 $\pm$  0.3 \\ 
G\_UM477          & 27 $\pm$  8 & 20 $\pm$ 10 & 31 $\pm$  8 &  6 $\pm$  3 & 15 $\pm$  7 & 47 $\pm$  7 & 21 $\pm$  7 &     7.1 $\pm$  0.7 \\ 
G\_UM103          &  6 $\pm$  5 & 10 $\pm$  6 & 45 $\pm$  8 & 15 $\pm$  7 & 23 $\pm$  8 & 17 $\pm$  6 & 38 $\pm$  8 &     7.6 $\pm$  1.0 \\ 
G\_NGC4507        &  3 $\pm$  2 &  6 $\pm$  4 &  4 $\pm$  3 & 26 $\pm$  9 & 61 $\pm$  9 &  9 $\pm$  4 & 87 $\pm$  4 &     7.1 $\pm$  2.3 \\ 
G\_NGC3281        &  1 $\pm$  1 &  4 $\pm$  2 &  5 $\pm$  3 & 39 $\pm$ 11 & 51 $\pm$ 11 &  5 $\pm$  3 & 90 $\pm$  4 &     7.3 $\pm$  2.9 \\ 
\hline
\end{tabular}
\end{centering}
\caption{Columns 2--6: Population vector in the
$(x_6,x_7,x_8,x_9,x_{10})$ description.  Columns 7, 4 and 8: $x_Y$,
$x_I$ and $x_O$ respectively. All $x$ components are in percentages of
the total flux at $\lambda_0 = 4020$ \AA. Column 9: mean starburst age
as defined in Section \ref{sec:Mean_age}. Objects starting with a G\_
are the spectral groups of Sample II. The three last entries in the
table correspond to Seyfert 2 systems.}
\label{tab:EPS_Results}
\end{table*}
%***TAB***TAB***TAB***TAB***TAB***TAB***TAB***TAB***TAB***TAB***TAB

%***FIG***FIG***FIG***FIG***FIG***FIG***FIG***FIG***FIG***FIG***FIG
\begin{figure*}
\resizebox{\textwidth}{!}{\includegraphics[18,400][592,718]{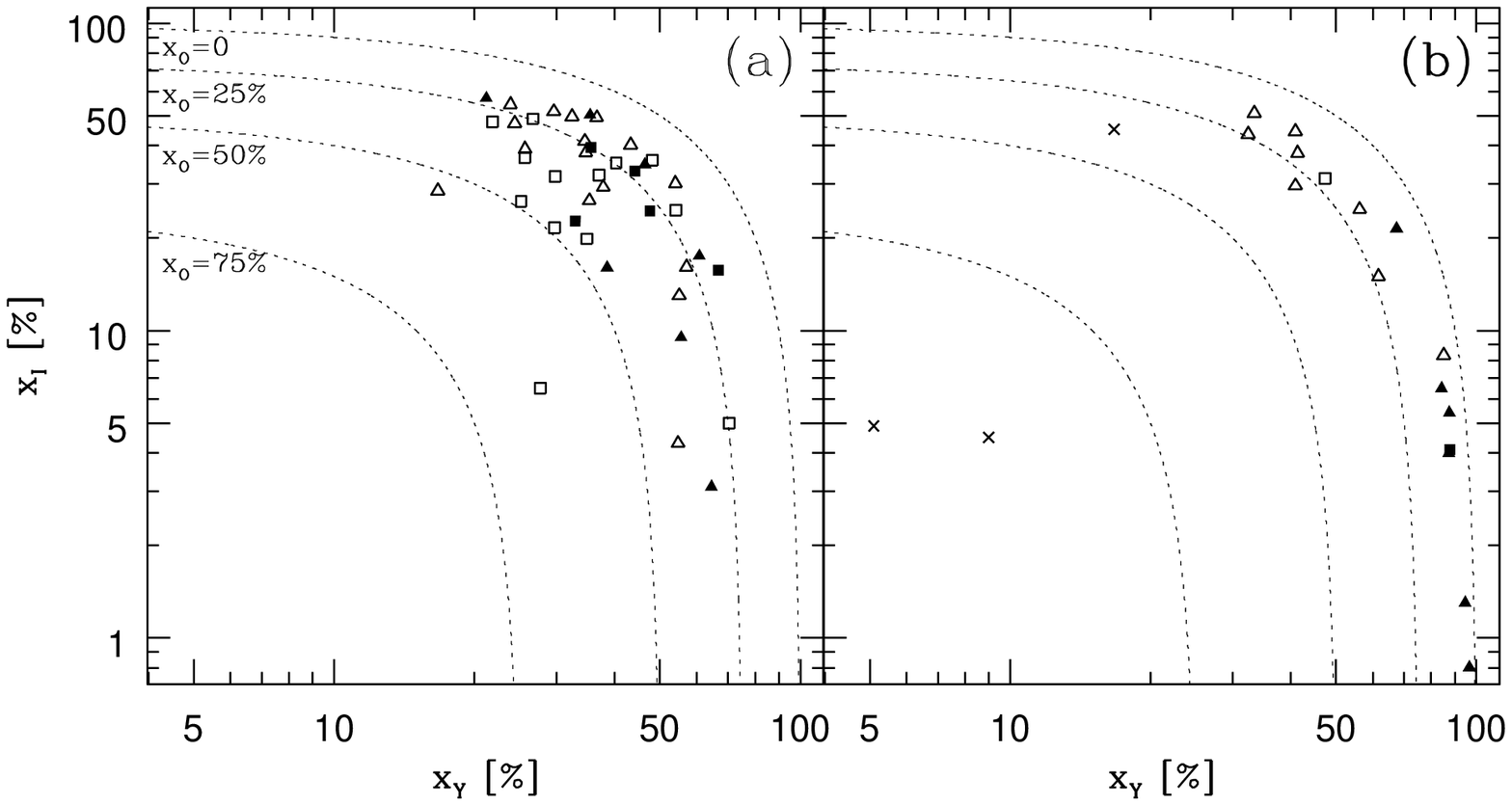}}
\caption{Results of the empirical population synthesis analysis for
Samples I (a) and II (b), condensed on an evolutionary diagram. The
horizontal axis $x_Y$ is the fraction of light at $\lambda_0 = 4020$
\AA\ due to stars in the $10^6$ and $10^7$ yr age bins, while the
fraction $x_I$ due to $10^8$ yr stars is plotted along the vertical
axis. A third perpendicular axis carries the contribution $x_O$ of
older populations ($\ge 10^9$ yr). Dotted lines indicate lines of
constant $x_O$, as labeled.  HII galaxies are plotted as triangles and
Starburst nuclei as squares. Filled symbols correspond to galaxies
with WR features. Crosses in panel (b) indicate Seyfert 2s.  Note that
the Starburst nucleus G\_Mrk710 and the HII galaxy G\_Tol1004\_296,
both of which show WR features, overlap at $(x_Y,x_I) = (0.87,0.04)$
in panel (b).}
\label{fig:evol1}
\end{figure*}
%***FIG***FIG***FIG***FIG***FIG***FIG***FIG***FIG***FIG***FIG***FIG

The results of the EPS analysis of galaxies in Samples I and II are
listed in Table \ref{tab:EPS_Results} both for the
$(x_6,x_7,x_8,x_9,x_{10})$ and the Young, Intermediate and Old
descriptions.  The normalization wavelength is $\lambda_0 = 4020$ \AA.
A very convenient way to present the results of the synthesis is to
project the $(x_Y,x_I,x_O)$ vector onto the $x_Y$-$x_I$ plane. This is
done in Figs.~\ref{fig:evol1}a and b for Samples I and II
respectively. Dotted lines in these plots mark lines of constant
$x_O$, computed from the $x_Y + x_I + x_O = 1$ condition. Note that
these are actually straight lines in the $x_Y$-$x_I$ plane, which
appear curved because of the logarithmic axis.

Galaxies from both samples define a smooth sequence from large $x_Y$
to large $x_I$, with not much spread in $x_O$, particularly for Sample
II. The larger spread seen in Fig.~\ref{fig:evol1}a is partly due to
the fact that the data for Sample I was collected through apertures
typically 2.6 times larger than for Sample II, and thus sample a more
heterogeneous mix of stellar populations. This interpretation is
supported by the fact that NGC 1510, which appears in both samples,
looks somewhat younger in Sample II (see Table \ref{tab:EPS_Results}).
Another source of scatter in Fig.~\ref{fig:evol1}a stems from the
large number of Starburst nuclei in Sample I. These systems,
represented by squares in both panels, live on galaxies with a
significant old stellar component, whose effect is to drag points
towards the bottom-left of the plot. For instance, the point at
$(x_Y,x_I,x_O) = (0.28,0.06,0.66)$ in Fig.~\ref{fig:evol1}a is NGC
5996, whose spectrum reveals a weak starburst immersed in an old
population (McQuade \etal 1995; Kennicutt 1992). HII galaxies, on the
other hand, are closer to ``pure starbursts''. In fact, they were once
thought to be truly young galaxies undergoing their first
star-formation episode (Searle \& Sargent 1972). Only recently it has
become clear that they too contain old stars (Telles \& Terlevich
1997; Schulte-Ladbeck \& Crone 1998; Raimann \etal 2000a). This
explains why Sample II, which is essentially an HII galaxy sample,
exhibits a more well defined sequence in Fig.~\ref{fig:evol1}, with
all non-AGN sources bracketed by the $x_O = 0$ and 30\% contours.

The three deviant crosses spoiling the Sample II sequence in
Fig.~\ref{fig:evol1}b are the Seyfert 2 groups, with their
predominantly old stellar populations (Raimann \etal 2000a).  G\_UM103
has a significant ``post-starburst'' component, reminiscent of more
evolved starburst + Seyfert 2 composite systems, while the other two
groups occupy a region characteristically populated by LINERs and
non-composite Seyfert 2's (Cid Fernandes \etal 2001b).

Since the location of a galaxy in Fig.~\ref{fig:evol1} reflects the
evolutionary state of its stellar population, we interpret the
distribution of objects in this diagram as an {\it evolutionary
sequence}, with the mean stellar age running counter-clockwise.

There are several reasons to interpret Fig.~\ref{fig:evol1} as an
evolutionary sequence. First, metal absorption lines become deeper and
galaxy colors become progressively redder as one moves from large
$x_Y$ to large $x_I$ along the sequence. In fact, the sequence defined
by Sample II follows very closely the blue to red (young to old)
spectral sequence in Figure 1 of Raimann \etal (2000a). Second, all
Sample II galaxies in which WR features have been detected (those
marked by filled symbols in Fig.~\ref{fig:evol1}b) are located in the
large $x_Y$ region of the diagram, consistent with the young burst
ages (a few Myr) implied by the mere presence of WR stars. Filled
symbols in Fig.~\ref{fig:evol1}a mark galaxies from Sample I which are
listed in the WR-galaxy catalog maintained by D. Schaerer
(webast.ast.obs-mip.fr/people/scharer). Their more even distribution,
as compared to Sample II, is due to the old population and aperture
effects discussed above and nicely illustrated by Meurer
(2000). Whereas the data analyzed here pertains to kpc-scales, spectra
used to classify a starburst as a WR-galaxy are usually obtained
through much narrower slits centered on the brightest cluster, thus
favoring the detection of young systems. Processing such spectra
through our EPS-machinery would surely move the filled points in
Fig.~\ref{fig:evol1}a towards younger ages. (Conversely, one would
expect that narrow slit observations of galaxies represented by empty
symbols in the bottom right of Fig.~\ref{fig:evol1}a, such as Mrk 357
and UGCA 410, have a good chance of revealing WR features.)  Finally,
galaxies located in the large $x_I$ zone in the top-left (such as NGC
1800 in Sample I and group G\_NGC3089 in Sample II) have spectra
typical of a ``post-starburst'' population, with pronounced high order
Balmer absorption lines typical of A stars (Gonz\'alez Delgado \etal
1999). It is important to remark that neither the presence of WR
features nor Balmer absorption lines were used in the EPS analysis,
and yet the EPS results are compatible with the information apported
by these observables.

\subsection{EPS analysis of theoretical galaxy spectra}

\label{sec:Bruzual}

%***FIG***FIG***FIG***FIG***FIG***FIG***FIG***FIG***FIG***FIG***FIG
\begin{figure*}
\resizebox{\textwidth}{!}{\includegraphics[18,420][592,718]{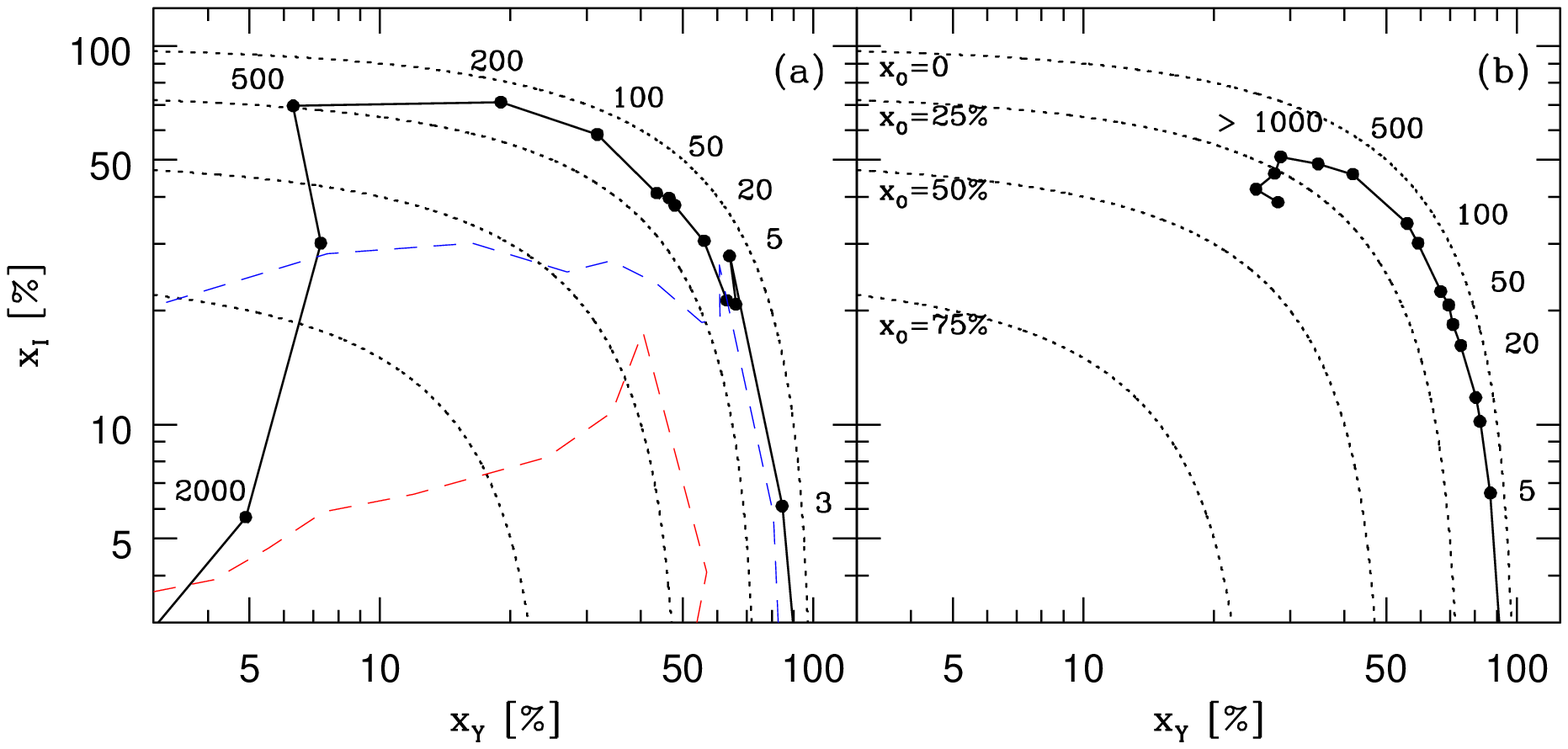}}
\caption{Evolution of GISSEL96 models on the $(x_Y,x_I,x_O)$ EPS
diagram, obtained processing the theoretical galaxies as the
data. Dotted lines indicate lines of constant $x_O$. (a) The solid
line shows results for an instantaneous burst. Numbers indicate the
model age in Myr (note that evolution proceeds counterclockwise).
Dashed lines correspond to instantaneous bursts on top of an old
population which at $t = 0$ accounts for $x_{Old}(0) = 10\%$ (top) and
50\% (bottom) of the flux at $\lambda_0 = 4020$ \AA.  (b) Models with
continuous star formation.}
\label{fig:YxI_BC}
\end{figure*}
%***FIG***FIG***FIG***FIG***FIG***FIG***FIG***FIG***FIG***FIG***FIG

%***FIG***FIG***FIG***FIG***FIG***FIG***FIG***FIG***FIG***FIG***FIG
\begin{figure*}
\resizebox{11cm}{!}{\includegraphics[18,144][592,640]{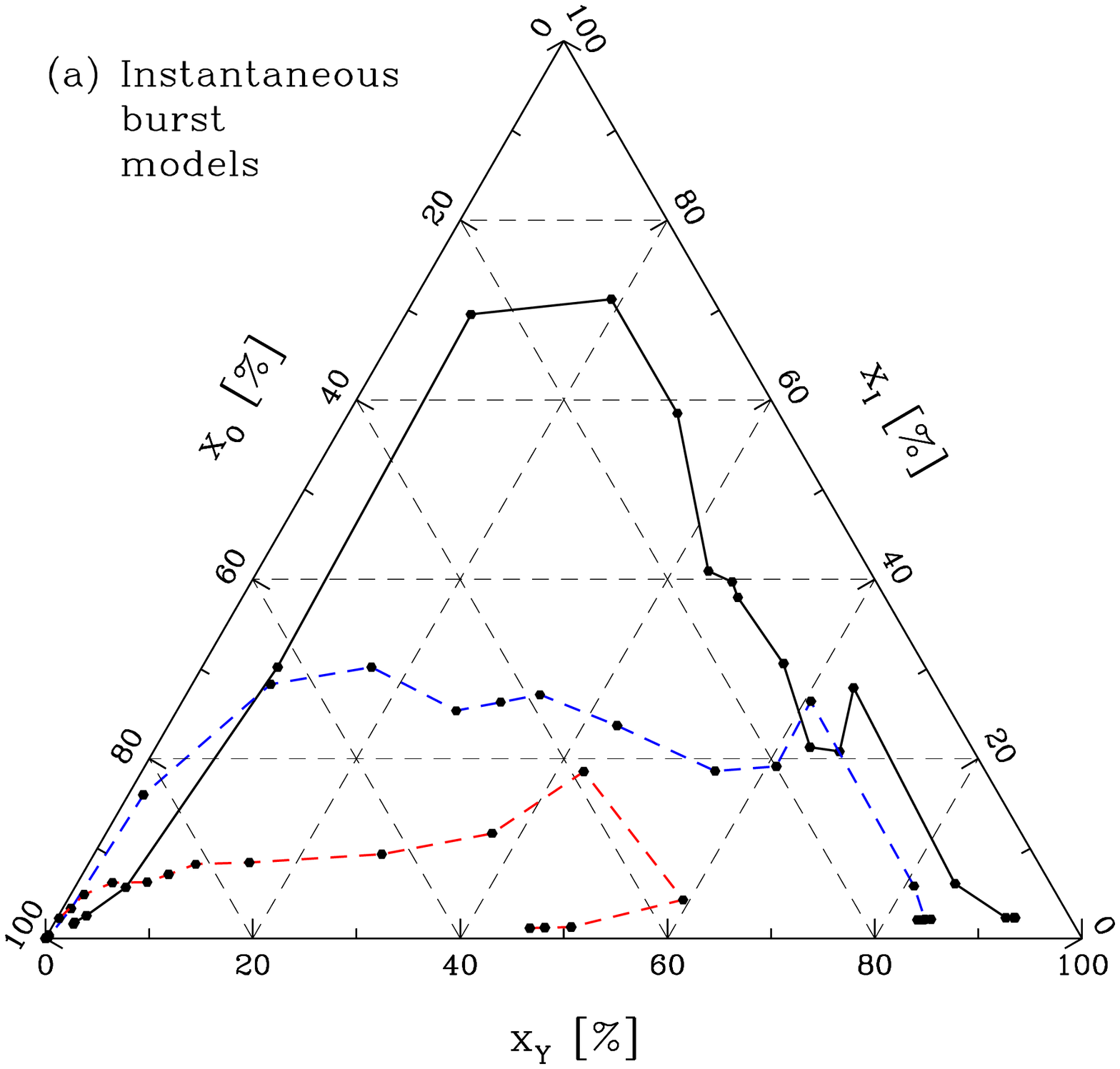}}
\resizebox{11cm}{!}{\includegraphics[18,144][592,650]{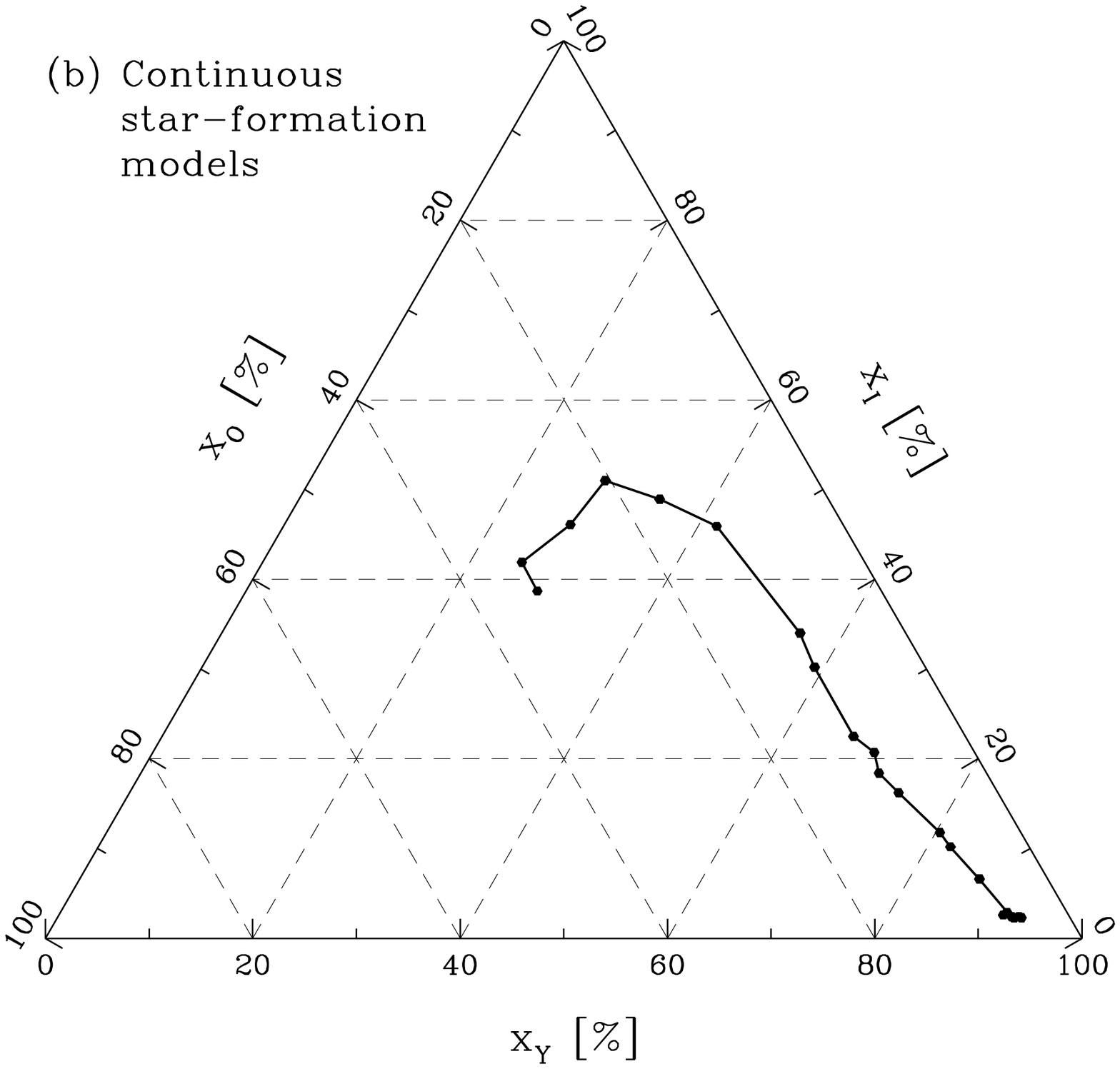}}
\caption{Evolution of GISSEL96 models in the $(x_Y,x_I,x_O)$ EPS-space
seen from a face on projection of the $x_Y + x_I + x_O = 1$ plane.
(a) Instantaneous burst models. The top curve corresponds to a pure
burst, while the middle and bottom curves correspond to bursts atop a
pre-existing old population with $x_{Old}(0) = 10$ and 50\%
respectively. (b) Continuous star formation models.}
\label{fig:tri_BC}
\end{figure*}
%***FIG***FIG***FIG***FIG***FIG***FIG***FIG***FIG***FIG***FIG***FIG

A straight-forward theoretical reason to interpret
Fig.~\ref{fig:evol1} as an evolutionary sequence is that,
schematicaly, a simple (i.e., coeval) stellar population moves on this
diagram from $(x_Y,x_I,x_O) = (1,0,0)$ at age $t = 0$ to $(0,1,0)$
after some $10^8$ yr and then to $(0,0,1)$ for ages $\ge 10^9$ yr. 

In order to follow this evolutionary path more closely we have carried
out an EPS analysis of {\it theoretical} galaxy spectra from GISSEL96,
the evolutionary synthesis code of Bruzual \& Charlot (1993). The
theoretical spectra were processed in exactly the same way as the real
spectra of Samples I and II. Instantaneous burst and continuous
star-formation models were computed for various ages between $t = 0$
and 15 Gyr, a Salpeter IMF between 0.1 and 125 $M_\odot$ and $Z =
Z_\odot$. GISSEL96 uses stellar tracks from the Padova group and
offers a choice of spectral libraries. We have chosen the one which
uses the atlas of Jacoby, Hunter \& Christian (1984) for the optical
range. The absorption features necessary for our EPS are clearly
defined with this library.  Spectral resolution was in fact the reason
we have chosen GISSEL96 over the Starburst99 code of Leitherer \etal
(1999), which is more taylored to study young stellar populations but
currently works with an optical library too coarse for EPS analysis.

\subsubsection{Burst models}

\label{sec:IB_models}

The results for an instantaneous burst are shown as a solid line in
the $(x_Y,x_I,x_O)$ evolutionary diagram of
Fig.~\ref{fig:YxI_BC}a. Labels next to selected points indicate the
model age in Myr. As expected, evolution proceeds from $x_Y$ to $x_I$
to $x_O$, such that a position on this diagram can be associated with
an age.  Fig.~\ref{fig:tri_BC}a provides an alternative representation
of this diagram, in which all 3 components are explicitly plotted in a
face-on projection of the plane containing the $(x_Y,x_I,x_O)$
vector. The idea for this projection was borrowed from similar plots
by Pelat (1997, 1998) and Moultaka \& Pelat (2001).

In principle one would expect all evolution prior to ages of $\sim
10^9$ yr to progress along the $x_O = 0$ contour, whereas in practice
the GISSEL96 models oscillate between $x_O = 10$ and 15\% for $t < 2
\times 10^8$ yr. Similarly, the value of $x_I$ starts to increase
before $t = 10^7$ yr, when all stars should still belong to the $x_Y$
age bin. These deviations occur due to the limited number of
observables used in the synthesis and because these contain
observational errors which broaden the likelihood-function of ${\bf
x}$ in a non-trivial way (Cid Fernandes \etal 2001a). As a result,
some of the true $x_Y$ proportion always spills over onto $x_I$ and
$x_O$, and so on.  Overall, however, these figures show an excellent
correspondance between empirical and evolutionary populations
synthesis calculations.

\begin{figure}
\resizebox{9cm}{!}{\includegraphics{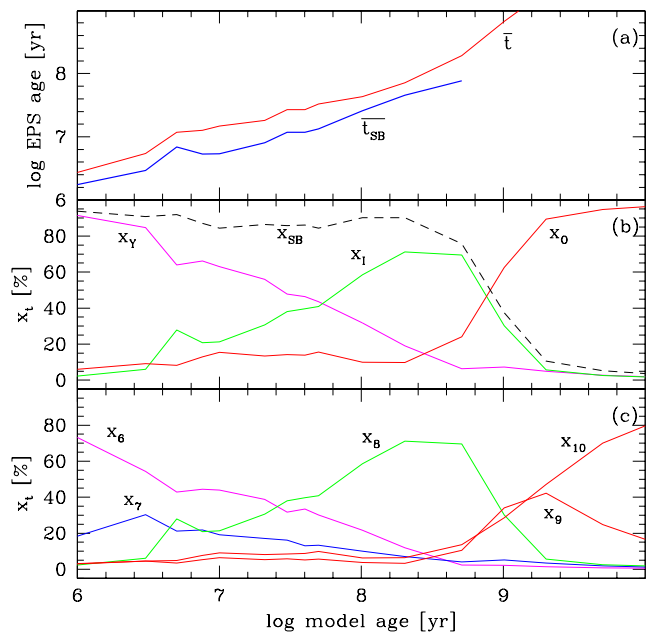}}
\caption{EPS parameters against model age for instantaneous burst
models. (a) Mean ages, (b) population vector ${\bf x}$ in the reduced
Young, intermediate and old description, and (c) in the 5 component
description. The $\ov{t}_{SB}$ curve in (a) is stopped at a few times
$10^8$ yr, after which it loses meaning for instantaneous bursts.}
\label{fig:calib_BC_IB}
\end{figure}

In Fig.~\ref{fig:calib_BC_IB}c we show the behavior of the 5 age
components $x_6$--$x_{10}$ as a function of the age of the GISSEL96
models. The population vector evolves smoothly with age, except for
the small kink a little short of $10^7$ yr due to the sudden
appearance of red supergiants (Charlot \& Bruzual 1991).  As expected,
$x_6$ peaks around $t = 10^6$ yr, $x_7$ peaks around $t = 10^7$ yr and
so on, but note how $x_7$ is less well defined than any other
component. The EPS decomposition tends to represent a $\sim 10^7$ yr
burst as a combination of $x_6$ and $x_8$ instead of a strong
$x_7$. Also, a non-negligible fraction of the $10^6$ yr component
spills from $x_6$ onto $x_7$ for $t \simless 10^6$ yr. A similar
effect occurs with $x_9$ and $x_{10}$ for $t \simgreat 10^9$ yr.

Such imprecisions in the mapping between evolutionary and empirical
population synthesis are largely suppressed in the coarser, but more
robust, $(x_Y,x_I,x_O)$ description, as shown in
Fig.~\ref{fig:calib_BC_IB}b. The figure also shows the evolution of
$x_{SB} = x_6 + x_7 + x_8$, which we hereafter treat as the
``starburst component'', representing the past $\sim 10^8$ yr of the
history of star formation in a galaxy. This is a more reasonable
definition for our purposes than using only the youngest, ionizing
population ($x_6$), since single burst models are not adequate to
describe kpc scale regions such as those sampled by the observations
of Samples I and II. Instead, the inner kpc of star-forming galaxies,
particularly Starburst nuclei, contains a collection of many
individual associations plus a field population with a spread in
age. The detailed studies by Lan\c{c}on \etal (2001) and Tremonti
\etal (2001) illustrate this point (see also Calzetti 1997; Legrand
\etal 2001). Such systems are frequently better represented by models
with multiple bursts or continuous star formation over $\sim 10^8$ yr
(Meurer 2000; Meurer 1995; Coziol, Barth \& Demers 1995; Coziol, Doyon
\& Demers 2001).

%***FIG***FIG***FIG***FIG***FIG***FIG***FIG***FIG***FIG***FIG***FIG
%***FIG***FIG***FIG***FIG***FIG***FIG***FIG***FIG***FIG***FIG***FIG

\subsubsection{Burst plus an underlying old population}

\label{sec:IB_plus_OLD_models}

Real galaxies have a mixture of stellar populations of different ages,
and galaxies in Samples I and II are no exception. The ongoing
star-formation which makes them classifiable as starburst systems is
observed atop an old ($\ge 10^9$ yr) stellar substrate formed in the
earlier history of the galaxy.  In our young, intermediate and old
description, the effect of this underlying population is to dilute the
values of $x_Y$ and $x_I$, which represent the recent history of
star-formation. As a result, an instantaneous burst occurring on top
of an old background does not follow the evolutionary sequence traced
by the solid line in Fig.~\ref{fig:YxI_BC}a.

Two quantities suffice to examine these diluting effects: the fraction
$x_{Old}(0)$ of the total $L_{Burst} + L_{Old}$ luminosity at the
start of the burst ($t=0$) which is due to old stars, and the function
$l(t) = L_{Burst}(t) / L_{Burst}(0)$, which describes the luminosity
evolution of the burst in units of its initial luminosity. Naturally,
all these quantities refer to the same wavelength, $\lambda_0$.  With
these definitions, and considering that $L_{Old}$ does not evolve
significantly on the time-scales of interest ($\le 10^8$ yr), it is
easy to show that

\begin{equation}
\label{eq:x_Old}
x_{Old}(t) = \frac{x_{Old}(0)}{x_{Burst}(0) l(t) + x_{Old}(0)}
\end{equation}

\ni where $x_{Burst}(0) = 1 - x_{Old}(0)$. We can now look at the
evolution of ${\bf x}$ presented above for a pure burst as
corresponding just to the $L_{Burst}(t)$ component, which at time $t$
accounts for only $x_{Burst}(t) = 1 - x_{Old}(t)$ of the total
luminosity of the system. This allows us to, with the help of
equation~\ref{eq:x_Old}, re-normalize the evolution of $(x_Y,x_I,x_O)$
to this new scale for any desired value of the {\it contrast
parameter} $x_{Old}(0)$.

Results for $x_{Old}(0) =10$ and 50\% are shown as dashed lines in
Figs.~\ref{fig:YxI_BC}a and \ref{fig:tri_BC}a. As the burst fades, the
evolutionary sequence bends over towards large $x_O$ quicker for
larger $x_{Old}(0)$, i.e., for smaller initial ratios of burst to
underlying old population power. Despite this effect, evolution still
proceeds in an orderly counterclockwise fashion.

\subsubsection{Continuous star-formation models}

\label{sec:CSF_models}

%***FIG***FIG***FIG***FIG***FIG***FIG***FIG***FIG***FIG***FIG***FIG
\begin{figure}
\resizebox{9cm}{!}{\includegraphics{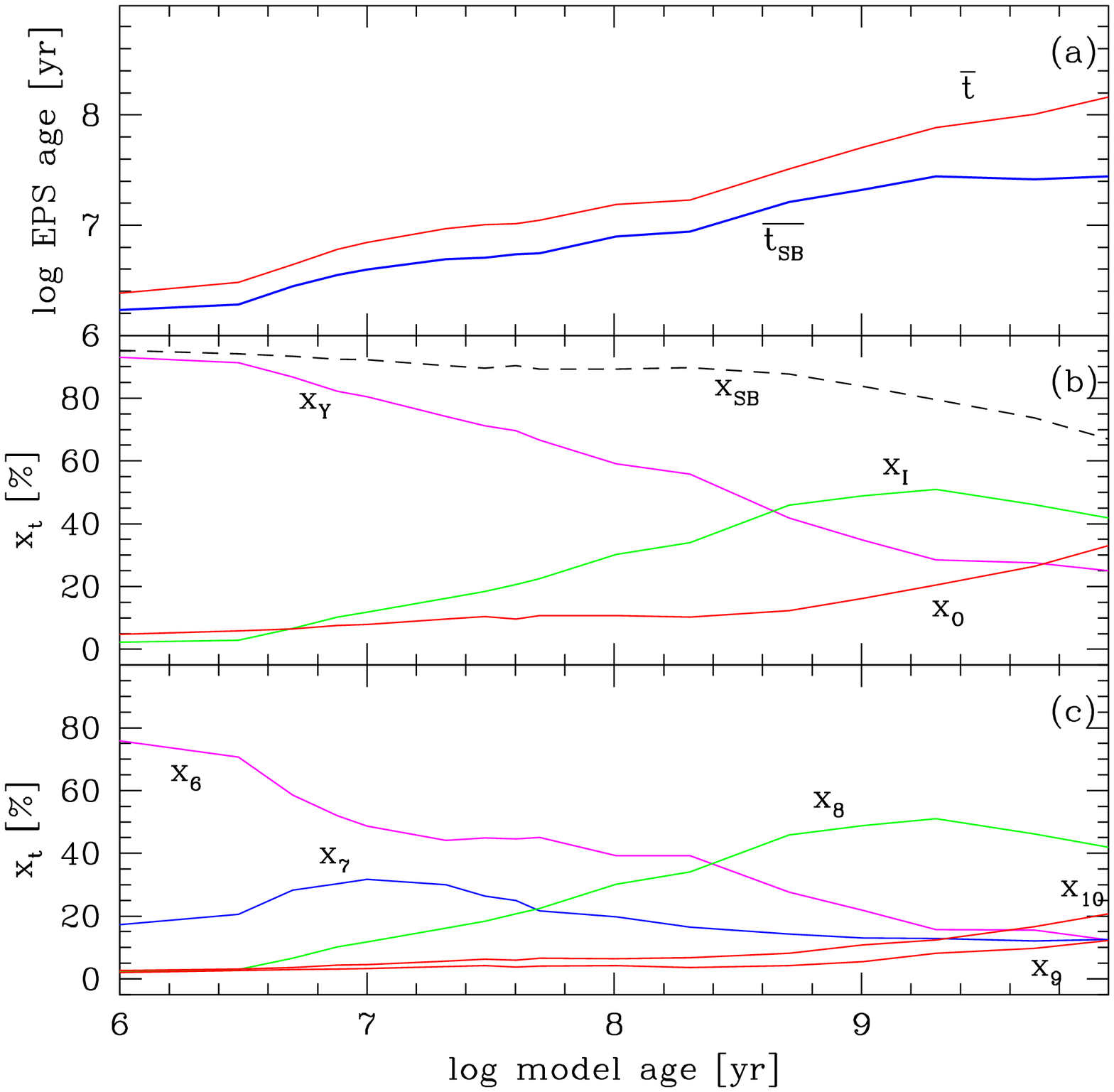}}
\caption{As Fig.~\ref{fig:calib_BC_IB} for models with continuous
star formation.}
\label{fig:calib_BC_CSF}
\end{figure}
%***FIG***FIG***FIG***FIG***FIG***FIG***FIG***FIG***FIG***FIG***FIG

Figs.~\ref{fig:YxI_BC}b and \ref{fig:tri_BC}b show the evolution of
$(x_Y,x_I,x_O)$ for GISSEL96 models with continuous
star-formation. Since in this regime there are always young stars, at
any given age $> 0$ the system looks younger than an instantaneous
burst model. At $t = 10^8$ yr, for instance, $(x_Y,x_I,x_O) \sim
(0.6,0.3,0.1)$ for the continuous star-formation models and
$(0.3,0.6,0.1)$ for an instantaneous burst. For $t > 10^9$ yr, ${\bf
x}$ converges to a region around $(0.3,0.4,0.3)$, instead of plunging
towards large $x_O$ as for instantaneous bursts.  Evolutionary
sequences for models with time-decaying star-formation rates (say,
exponentially or an ``extended burst'' step function), would define
curves intermediary between those traced in Figs.~\ref{fig:YxI_BC}a
and b.

Since the luminosity increases without bounds for continuous star
formation, any underlying old population is quickly outshone by the
new stars. We therefore do not present dilution curves such as those
computed for an instantaneous burst, since, except for the earliest
ages, they are practically identical to the undiluted curves in
Figs.~\ref{fig:YxI_BC}b and \ref{fig:tri_BC}b. The evolution of the
population vector for continuous star-formation models is shown in
Figs.~\ref{fig:calib_BC_CSF}b and c.  As expected, the curves are
smoother, and each component lives longer than for an instantaneous
burst (Fig.~\ref{fig:calib_BC_IB}).

\subsection{Mean stellar age}

\label{sec:Mean_age}

The experiments above demonstrate that the evolutionary state of a
starburst can be assessed by its location on the $(x_Y,x_I,x_O)$
diagram. In order to translate this location into a number which
quantifies the ``evolutionary state'' one could use, for instance, the
angle $\theta = \tan^{-1} x_I/x_Y$, which increases as a burst
evolves. Alternatively, we may use ${\bf x}$ to compute the {\it mean
age} $\ov{t}$ of the stellar population.  Since stellar populations
evolve in a non-linear way, it makes more sense to define $\ov{t}$
from the mean $\log t$ among the populations represented by the base:

\begin{equation}
\label{eq:Def_log_t_flux}
\log \ov{t}(\lambda_0) \equiv \sum x_i(\lambda_0) \log t_i
\end{equation}

The dependence of ${\bf x}$ on the normalization wavelength is
explicitly written in this equation to emphasize that this definition
of $\ov{t}$ is $\lambda$-dependent.  This happens because the $x_i$'s
are flux fractions at $\lambda_0$, so $\ov{t}(\lambda_0)$ is a {\it
flux-weighted mean age}. A $\lambda$-dependent age makes observational
sense for the simple reason that young stars are bluer than old stars,
which makes $\ov{t}$ an increasing function of $\lambda$.  Though the
value of $\ov{t}$ depends on $\lambda_0$, the evolutionary sequence
traced by this index independs on the choice of normalization, and so
it can be used to {\it rank} populations on different evolutionary
states.

For our 5-ages base ($t_i = 10^6$, $10^7$, $10^8$, $10^9$ and
$10^{10}$ yr), $\ov{t}(\lambda_0)$ becomes (in yr)

\begin{equation}
\label{eq:Def_log_t_flux_5ages}
\log \ov{t}(\lambda_0) \equiv 
6 x_6 + 7 x_7 + 8 x_8 + 9 x_9 + 10 x_{10}
\end{equation}

Since we are primarily interested in quantifying the evolutionary
stage of populations associated with the most recent star-formation in
starburst systems, it is interesting to consider a definition of
$\ov{t}$ which removes the diluting effects of an underlying old
stellar population. This can be done re-normalizing $x_6 + x_7 + x_8$
to 1, which yields the following definition for the {\it mean
starburst age} (also in yr):

\begin{equation}
\label{eq:Def_log_tSB_flux}
\log \ov{t}_{SB}(\lambda_0) \equiv 
\frac{6 x_6 + 7 x_7 + 8 x_8}{x_6 + x_7 + x_8}
\end{equation}

\noindent Note that by construction $6 \le \log \ov{t} \le 10$ and
$6 \le \log \ov{t}_{SB} \le 8$.

The solid lines in Fig.~\ref{fig:calib_BC_IB}a compare our $\ov{t}$
and $\ov{t}_{SB}$ EPS-based age indices for $\lambda_0 = 4020$ \AA\
with the corresponding age of the GISSEL96 models for an instantaneous
burst.  Despite some minor oscillations, these two indices bear a
$\sim$ one-to-one relation with the theoretical age.
Fig.~\ref{fig:calib_BC_CSF}a presents these same indices but for the
continuous star-formation models.  As expected, $\ov{t}$ and
$\ov{t}_{SB}$ evolve more slowly than for an instantaneous burst,
but they still increase steadily with the model age.  Since $\ov{t}$
and $\ov{t}_{SB}$ are entirely obtained from a few easily measurable
quantities, this result encourages their use as {\it empirical clocks}
for stellar populations.

%\subsubsection{The use of $\ov{t}_{SB}$ as a clock for starbursts}
%\label{sec:mean_age_discussion}

As it is clear from its very definition, due to the coarse
age-resolution of the base, our mean age index $\ov{t}_{SB}$ is not
meant to be used as a fine-graded chronometer of starbursts. Yet, the
above experiments with theoretical spectra clearly show that
$\ov{t}_{SB}$ provides a useful way to {\it rank} galaxies according
to the age of the dominant population among the multiple generations
of stars formed in the recent history of star-formation.  This
definition is particularly well suited to describe spectra which are
integrated over large regions and hence average over many such
generations, Although all galaxies discussed here contain populations
younger than $10^7$ yr, which power their emission line spectrum, this
ongoing star-formation may be less intense than in the recent past
($10^7$--$10^8$ yr), such that the young generations live among an
older, non-ionizing starburst population. In this case, one expects to
find significant $x_7$ and $x_8$ components, and thus $\ov{t}_{SB} >
10^7$ yr. Conversely, if the current star-formation is more vigorous
than in the past, mean ages of less than $10^7$ yr are expected. It is
in this context of starbursts extended over a period of up to $\sim
10^8$ yr that we envisage $\ov{t}_{SB}$ and the $(x_Y,x_I,x_O)$
diagram as useful tracers of evolution.

In principle, a base with a finer age resolution, including elements
intermediate between $x_6$, $x_7$ and $x_8$, could yield a more
detailed description of the evolution of starbursts.  In practice,
however, these elements would be well approximated by linear
combinations of the existing base elements unless new observables were
introduced in the synthesis process. For this reason, we opted to
perform our EPS analysis with the base and observables described in
\S\ref{sec:Synthesis_Method}, whose pros and cons have already been
fully exploited in our previous investigations (Cid Fernandes \etal
2001a,b; Schmitt \etal 1999). Furthermore, as we shall soon see,
this relatively coarse description is well suited to our present
purposes.

\section{The evolution of emission line properties}

\label{sec:Emission_Line_Properties}

Emission lines in star-forming galaxies are umbilicaly linked to
their young stellar population, whose massive, hot stars photoionize
the surrounding gas. Also, in non-instantaneous starbursts the
continuum carries a large contribution of stars borne before the
current generation of ionizing stars, thus affecting emission line
equivalent widths. In this section, we combine the tools to measure
the evolution of starbursts developed in Section \ref{sec:Synthesis}
with the emission line data compiled in Section \ref{sec:Data} to
investigate whether the emission line properties do indeed evolve
along with the burst.

\subsection{The equivalent width of H$\beta$}

\label{sec:W_Hb}

%***FIG***FIG***FIG***FIG***FIG***FIG***FIG***FIG***FIG***FIG***FIG
\begin{figure}
\resizebox{9cm}{!}{\includegraphics{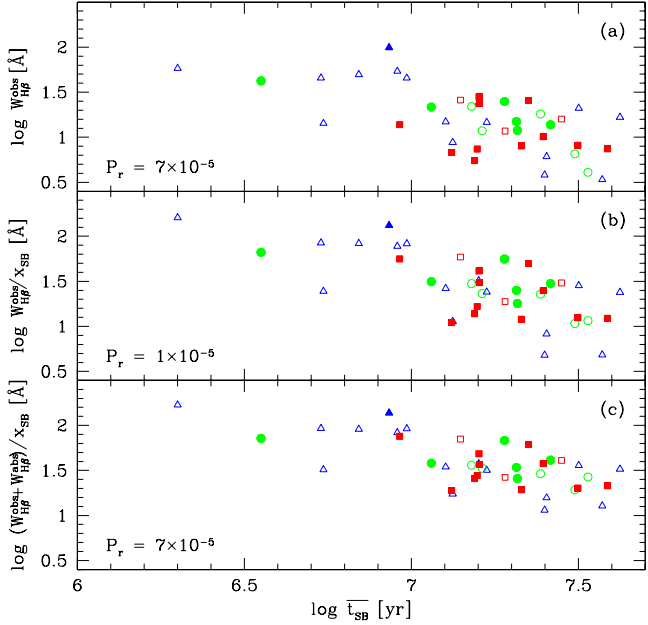}}
\caption{Equivalent width of H$\beta$ against the EPS-defined mean
starburst age $\ov{t}_{SB}$ for Sample I.  Panel (a) shows the
observed equivalent widths ($W_{H\beta}^{obs}$). In (b) the
contribution of old stars to the continuum under H$\beta$ is removed
dividing $W_{H\beta}^{obs}$ by $x_{SB}(4861)$. In panel (c)
$W_{H\beta}$ is further corrected for the presence of an absorption
component. Triangles, circles and squares correspond to gas
metallicity ranges of (O/H) $< 0.4$, 0.4--0.6 and $> 0.6$
(O/H)$_\odot$ respectively. Open and filled symbols are used to
distinguish HII galaxies from Starburst nuclei.  The $P_r$ values are
the probabilities of no correlation in a Spearman test, small values
indicating significant correlations.}
\label{fig:W_Hb_x_age_SBMQ}
\end{figure}
%***FIG***FIG***FIG***FIG***FIG***FIG***FIG***FIG***FIG***FIG***FIG

%***FIG***FIG***FIG***FIG***FIG***FIG***FIG***FIG***FIG***FIG***FIG
\begin{figure}
\resizebox{9cm}{!}{\includegraphics{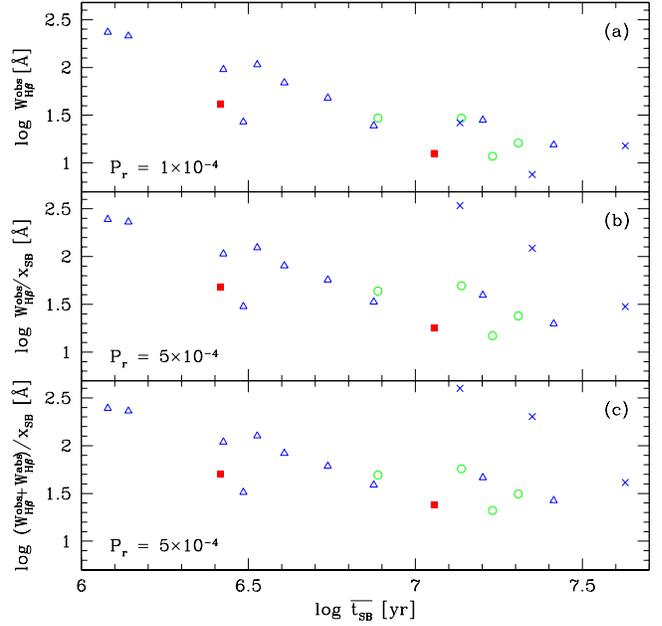}}
\caption{As Fig.~\ref{fig:W_Hb_x_age_SBMQ} but for Sample II. Crosses
indicate Seyfert 2's, which were excluded from the correlation
analysis.}
\label{fig:W_Hb_x_age_HIIgal}
\end{figure}
%***FIG***FIG***FIG***FIG***FIG***FIG***FIG***FIG***FIG***FIG***FIG

As a burst ages, its most massive stars are the first to die,
resulting in a steady decline of the ionizing photon flux and hence on
the luminosity of recombination lines such as H$\beta$. The stellar
continuum $C_{H\beta}$ underneath H$\beta$ also decreases, but more
slowly than $L_{H\beta}(t)$, since it carries a significant
contribution from longer-lived, non-ionizing, lower mass stars. As a
result $W_{H\beta} = L_{H\beta} / C_{H\beta}$ decreases as the burst
evolves, as first discussed by Dottori (1981) and confirmed by
evolutionary synthesis calculations. $W_{H\beta}$ is therefore an {\it
age indicator}, and it is frequently used as such in studies of
star-forming systems (e.g, Stasi\'nska \etal 2001). However, an
empirical confirmation of the $W_{H\beta}(t)$ prediction requires an
{\it independent} measure of the starburst age.

In Figs.~\ref{fig:W_Hb_x_age_SBMQ}a and \ref{fig:W_Hb_x_age_HIIgal}a
we carry out this test with galaxies from Samples I and II
respectively, using our EPS-based index $\ov{t}_{SB}$ as a clock for
the starburst.  The anti-correlation between $W_{H\beta}$ and
$\ov{t}_{SB}$ is evident for both samples, thus confirming that
$W_{H\beta}$ decreases with time.  The probability $P_r$ of no
correlation in a Spearman's rank test is just $7 \times 10^{-5}$ for
Sample I and $10^{-4}$ for Sample II, indicating a very high
statistical significance.  We emphasize that $W_{H\beta}$ and
$\ov{t}_{SB}$ are determined from {\it completely independent}
measurements, which only highlights the significance of this result.

The different symbols in Figs.~\ref{fig:W_Hb_x_age_SBMQ},
\ref{fig:W_Hb_x_age_HIIgal} and all subsequent plots represent three
gas metallicity ranges, triangles, circles and squares corresponding
to (O/H) $< 0.4$, 0.4 to 0.6 and $> 0.6$ (O/H)$_\odot$
respectively. Open and filled symbols are used to distinguish HII
galaxies from Starburst nuclei. We postpone a discussion of the
effects of $Z$ and activity class to Sections \ref{sec:Excitation} and
\ref{sec:Metallicity}, which explore emission line properties more
directly affected by these variables.

\subsubsection{Corrections to $W_{H\beta}$}

\label{sec:W_Hb_corrections}

The values of $W_{H\beta}$ in Figs.~\ref{fig:W_Hb_x_age_SBMQ}a and
\ref{fig:W_Hb_x_age_HIIgal}a are the raw measurements
($W^{obs}_{H\beta}$). At least two corrections have to be considered,
both of which increase $W_{H\beta}$.

(1) The continuum under H$\beta$ carries a contribution from an old
stellar population which dilutes $W_{H\beta}$ with respect to the
value it would have in a pure starburst. Our EPS analysis provides a
natural way of correcting for this effect, which is a major source of
concern in studies which use $W_{H\beta}$ as an age-indicator
(Stasi\'nska \etal 2001 and references therein).  In order to isolate
the contribution of the starburst to $C_{H\beta}$ it suffices to
multiply it by $x_{SB}(4861)$, the starburst component defined in
Section \ref{sec:IB_models} but renormalized to $\lambda = 4861$
\AA. The corrected $W_{H\beta}$ is thus simply $W^{obs}_{H\beta} /
x_{SB}(4861)$. Figs.~\ref{fig:W_Hb_x_age_SBMQ}b and
\ref{fig:W_Hb_x_age_HIIgal}b show the dilution-corrected evolution of
$W_{H\beta}$.

The effects of this correction are largest for two of the Seyfert 2
groups in Sample II, which move significantly above the sequence in
Fig.~\ref{fig:W_Hb_x_age_HIIgal}b because of their bulge dominated
optical continuum (large $x_O$). The effect is not so large for
G\_UM103, which, as already discussed, resembles a starburst + Seyfert
2 composite. The fact that the observed values of $W_{H\beta}$ in
Seyfert 2's are in general smaller than those in starburst systems is
purely due to this dilution effect.  As explained by Cid Fernandes
\etal (2001b), Seyfert 2's {\it ought} to have intrinsically larger
$W_{H\beta}$ than starbursts, as we obtain with our EPS-based dilution
correction.

Among the star-forming galaxies in Samples I and II, this correction
typically increases $W_{H\beta}$ by $\sim 50\%$ (in good agreement
with the corrections inferred by Mas-Hesse \& Kunth 1999 on similar
objects), but it reachs more than a factor of 2 in some cases. The
correction is somewhat smaller for Sample II, partly because it
contains intrinsically younger systems and partly because of its
smaller apertures, which reduces ``contamination'' by an extended old
stellar population.  The dilution correction improves the
$W_{H\beta}$-$\ov{t}_{SB}$ correlation for Sample I and degrades the
one for Sample II, while for the combined sample the statistical
significance remains unchanged at $P_r = 5 \times 10^{-10}$.

(2) A second correction to be considered is that due to the presence
of an absorption component hidden underneath the H$\beta$
emission. This component is present in our spectral base with
strengths of up to $W^{abs}_{H\beta} = 8$ \AA, achieved for $10^8$ yr
populations.  We have used the population vector ${\bf x}$ obtained in
the synthesis to compute the expected value of $W^{abs}_{H\beta}$,
typically 3--5 \AA.  Adding $W^{abs}_{H\beta}$ to $W^{obs}_{H\beta}$
yields a corrected emission $W_{H\beta}$.  The combined dilution and
absorption corrected values of $W_{H\beta}$ are shown in
Figs.~\ref{fig:W_Hb_x_age_SBMQ}c and \ref{fig:W_Hb_x_age_HIIgal}c.

The absorption correction is only significant for galaxies
with weak H$\beta$ emission, such as those bellow $W^{obs}_{H\beta}
\sim 10$ \AA\ at the bottom right of
Fig.~\ref{fig:W_Hb_x_age_SBMQ}a. For these systems one has to look at
the absorption-corrected values of $W_{H\beta}$ as uncertain by as
much as a factor of 2.  The correction is negligible for most galaxies
in Sample II, which, due to its objective prism selection criterium,
contains more strong lined objects than Sample I. Indeed, the mean
$W^{obs}_{H\beta}$ is 55 \AA\ for Sample II but just 22 \AA\ for
Sample I. This is also why, as a whole, Sample II contains a higher
proportion of young starbursts.

The absorption correction degrades the $W_{H\beta}$-$\ov{t}_{SB}$
correlation for Sample I slightly.  For the combined Sample I + II
data the $P_r$ value increases from $5 \times 10^{-10}$ to $3 \times
10^{-8}$, which is still significant at the 5-$\sigma$ level. We thus
see that these `1$^{\rm st}$ order corrections' introduce very little
scatter, and do not alter our conclusion that $W_{H\beta}$ does indeed
evolve along with the stars that make up a starburst. One can also
look at this result the other way around, and conclude that the fact
that the $W_{H\beta}$ versus $\ov{t}_{SB}$ diagram behaves as expected
proves the usefulness of our EPS-based evolutionary index
$\ov{t}_{SB}$, with the advantage that it is immune to the dilution
and absorption effects which plague $W_{H\beta}$ and other
emission-line based age indicators.

We note in passing that these corrections alone are enough to bring
the values of $W_{H\beta}$ within the range spanned by evolutionary
synthesis calculations such as those by Leitherer \etal (1999),
whereas, as it has long been known, the raw observed values fall
bellow such predictions (Bressolin, Kennicutt \& Garnett 1999 and
references therein). Differential extinction, with line emitting
regions being more reddened than the stellar continuum (Calzetti,
Kinney \& Storchi-Bergmann 1994) and leakage of ionizing photons out
of the HII regions associated with the starburst are further examples
of processes that act in the sense of reducing $W_{H\beta}$.  We thus
concur with Raimann \etal (2000b) and Stasi\'nska \etal (2001) in that
the apparent mismatch between theoretical and observed values of
$W_{H\beta}$ bears no fundamental physical significance. In fact, the
combined effects of differential extinction, leakage and the
uncertainties in the dilution and absorption corrections are probably
responsible for most of the vertical scatter in
Figs.~\ref{fig:W_Hb_x_age_SBMQ}c and \ref{fig:W_Hb_x_age_HIIgal}c.

\subsubsection{Comparison with models}

\label{sec:W_Hb_x_comparison_with_SB99}

%***FIG***FIG***FIG***FIG***FIG***FIG***FIG***FIG***FIG***FIG***FIG
\begin{figure}
\resizebox{9cm}{!}{\includegraphics{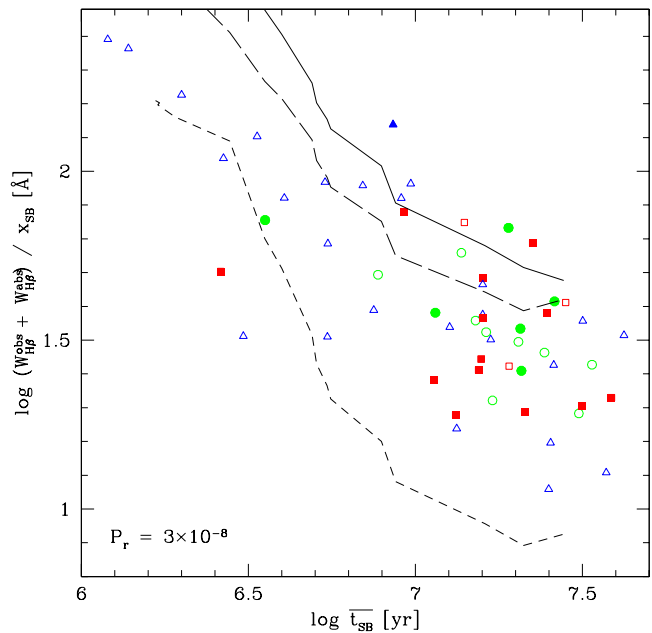}}
\caption{Observed and predicted evolution of $W_{H\beta}$. Data points
are as in Figs.~\ref{fig:W_Hb_x_age_SBMQ}c and
\ref{fig:W_Hb_x_age_HIIgal}c. Lines correspond to theoretical
$W_{H\beta}(t)$ curves with $t$ converted onto the EPS-based
$\ov{t}_{SB}$ scale. {\it Solid line:} GISSEL96 models with a
Salpeter IMF between 0.1 and 125 M$_\odot$.  {\it Dashed lines:}
Starburst99 models with a Salpeter IMF with M$_{upp} = 100$ M$_\odot$
(long dashes) and 30 M$_\odot$ (short dashes).}
\label{fig:W_Hb_x_models}
\end{figure}
%***FIG***FIG***FIG***FIG***FIG***FIG***FIG***FIG***FIG***FIG***FIG

A quantitative comparison of the predicted evolution of $W_{H\beta}$
with that detected in Figs.~\ref{fig:W_Hb_x_age_SBMQ} and
\ref{fig:W_Hb_x_age_HIIgal} demands processing model spectra through
the same EPS machinery used to analyse the data. This is necessary to
translate model ages ($t$) onto our $\ov{t}_{SB}$ age scale. We have
used the evolution of the rate of ionizing photons $N(H^0)$ and the
continuum under H$\beta$ predicted by GISSEL96 to compute
$W_{H\beta}(t)$ for the same models synthesized in Section
\ref{sec:Bruzual}, for which the $t \rightarrow \ov{t}_{SB}$
conversion is shown in Figs.~\ref{fig:calib_BC_IB}a and
\ref{fig:calib_BC_CSF}a.

The result for the continuous star formation models is shown as a
solid line in Fig.~\ref{fig:W_Hb_x_models}, over-plotted onto the data
points from Samples I and II (Figs.~\ref{fig:W_Hb_x_age_SBMQ}c and
\ref{fig:W_Hb_x_age_HIIgal}c).  The dashed curves in this plot are the
Starburst99 predictions (Leitherer \etal 1999) for solar metallicity,
constant star formation models with a Salpeter IMF up to M$_{upp} =
100$ and 30 M$_\odot$. These latter curves are drawn with the $t
\rightarrow \ov{t}_{SB}$ conversion obtained for GISSEL96, since, as
already explained, it is not currently possible to do an EPS analysis
with Starburst99 due to its poor spectral resolution optical
libraries. Instantaneous burst models (not shown for clarity) follow
roughly the same curves up to $\ov{t}_{SB} \sim 10^{6.8}$ yr and then
plunge vertically, signaling the end of the ionizing phase of the
cluster.

The rate at which $W_{H\beta}$ evolves is similar for models and data.
Furthermore, practically all data points are bracketed by the models
shown!  Given the already discussed caveats affecting both axis of
this figure, it would be premature to use this result to draw any
conclusion about, say, the IMF in starbursts.  The point to emphasize
here is that, to our knowledge, this is the first time that predicted
and observed values of $W_{H\beta}$ are plotted against an age axis,
and it is gratifying to see a good agreement between data and models.

\subsection{The equivalent width of [OIII]}

\label{sec:W_O3}

%***FIG***FIG***FIG***FIG***FIG***FIG***FIG***FIG***FIG***FIG***FIG
\begin{figure}
\resizebox{9cm}{!}{\includegraphics{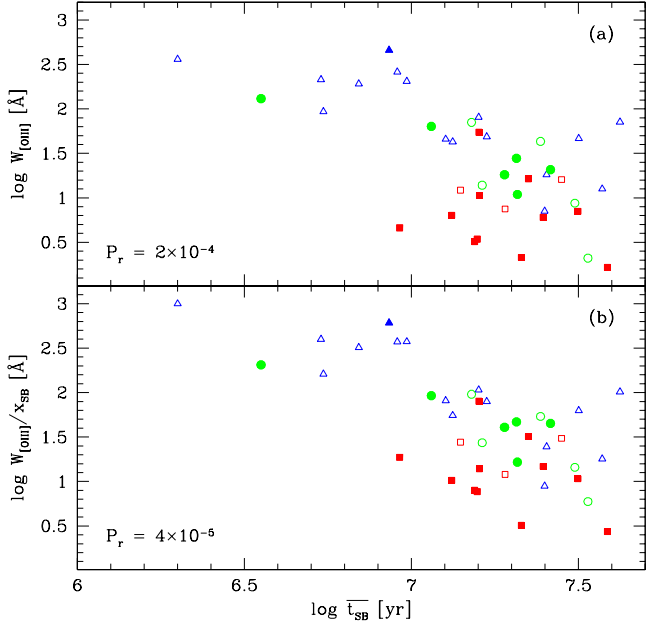}}
\caption{Equivalent width of [OIII]$\lambda$5007 against the mean
starburst age $\ov{t}_{SB}$ for Sample I.  Panel (a) shows the
observed equivalent widths, whereas in (b) the dilution of
$W_{[OIII]}$ by the underlying old stellar population is removed.
Symbols as in Fig.~\ref{fig:W_Hb_x_age_SBMQ}.}
\label{fig:W_O3_x_age_SBMQ}
\end{figure}
%***FIG***FIG***FIG***FIG***FIG***FIG***FIG***FIG***FIG***FIG***FIG

The popularity of $W_{H\beta}$ as an age indicator stems mostly from
its insensitivity to nebular conditions such as density, temperature
and metallicity. Yet, the detailed photoionization models for evolving
starbursts by Stasi\'nska \& Leitherer (1996) show that the equivalent
width of [OIII] is also a powerful chronometer of starbursts for
metallicities below solar, as is the case for most of the galaxies
studied here. We therefore explore the behavior of $W_{[OIII]}$
against our empirical age index $\ov{t}_{SB}$.

The results for Sample I are shown in
Fig.~\ref{fig:W_O3_x_age_SBMQ}. An anti-correlation is clearly
present. The relation gets even stronger after correcting $W_{[OIII]}$
for the dilution by an underlying population
(Fig.~\ref{fig:W_O3_x_age_SBMQ}b). Triangles and circles, which
correspond to the two lower $Z$ bins, trace rather well defined
sequences in Fig.~\ref{fig:W_O3_x_age_SBMQ}b, but the more metal rich
galaxies (plotted as squares) present a more scattered distribution.
A plausible explanation for this larger spread is that, as discussed
by Stasi\'nska \& Leitherer (1996), $W_{[OIII]}$ ceases to be a
decreasing function of age as $Z$ approaches $Z_\odot$.
Fig.~\ref{fig:W_O3_x_age_SBMQ}b also shows that the higher $Z$
galaxies tend to have older starbursts, an effect which is further
discussed below.

\subsection{Gas excitation}

\label{sec:Excitation}

%***FIG***FIG***FIG***FIG***FIG***FIG***FIG***FIG***FIG***FIG***FIG
\begin{figure}
\resizebox{9cm}{!}{\includegraphics{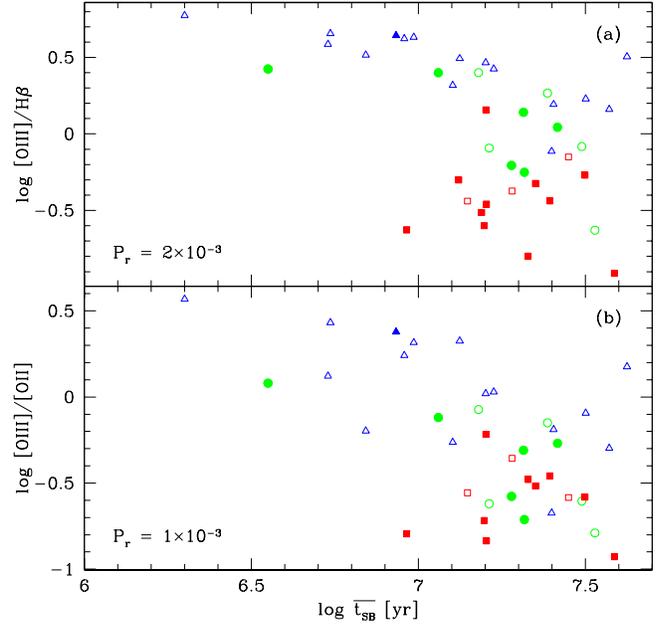}}
\caption{The evolution of gas excitation for Sample I galaxies, as
measured by (a) [OIII]/H$\beta$ and (b) [OIII]/[OII].  Reddening and
H$\beta$ absorption corrections were applied. Symbols as in
Fig.~\ref{fig:W_Hb_x_age_SBMQ}.}
\label{fig:Excitation_SBMQ}
\end{figure}
%***FIG***FIG***FIG***FIG***FIG***FIG***FIG***FIG***FIG***FIG***FIG

%***FIG***FIG***FIG***FIG***FIG***FIG***FIG***FIG***FIG***FIG***FIG
\begin{figure}
\resizebox{9cm}{!}{\includegraphics{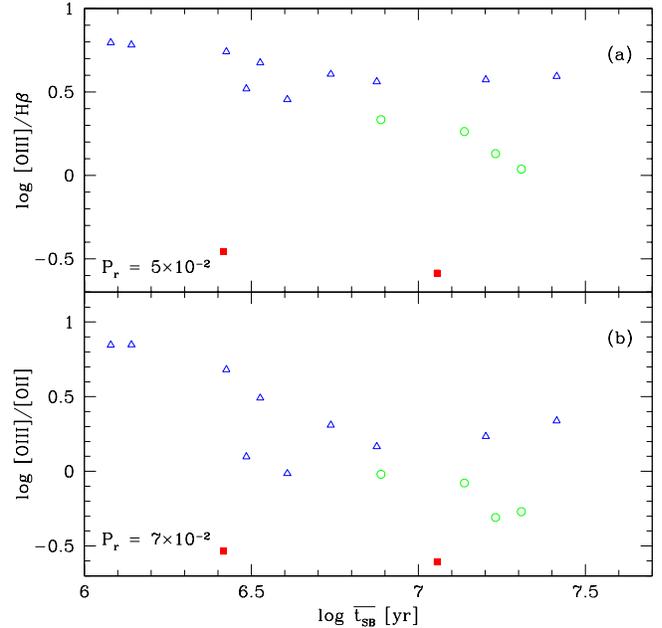}}
\caption{As Fig.~\ref{fig:Excitation_SBMQ} but for non-Seyfert
galaxies in Sample II.}
\label{fig:Excitation_Raimann}
\end{figure}
%***FIG***FIG***FIG***FIG***FIG***FIG***FIG***FIG***FIG***FIG***FIG

Another prediction of evolutionary synthesis plus photoionization
calculations is that the gas excitation decreases with time, because
the ratio of ionizing photons per gas particle (the ``ionization
parameter'') decreases and the ionizing spectrum softens as the hotter
stars die (e.g., Copetti \etal 1986; Cid Fernandes \etal 1992;
Garc\'{\i}a-Vargas \etal 1995; Stasi\'nska \& Leitherer 1996).  In
Figs.~\ref{fig:Excitation_SBMQ} and \ref{fig:Excitation_Raimann} we
explore the evolution of two line ratios in order to address this
issue.

Unlike for emission line equivalent widths, particularly $W_{H\beta}$,
metallicity plays a key role in defining ratios involving forbidden
lines because of its influence on the gas temperature. The three $Z$
intervals represented by different symbols in these figures help to
disentangle the effects of evolution and $Z$. Since HII galaxies are
less chemically evolved than Starburst nuclei, differences in
metallicity should also become apparent distinguishing objects by
their activity class. This can be readily seen in
Figs.~\ref{fig:W_Hb_x_age_SBMQ}--\ref{fig:Metallicity}.  Most HII
galaxies (open symbols) are in the low $Z$ bin (triangles), while most
Starburst nuclei (filled symbols) are in the high $Z$ bin
(squares). Sample I contains only one Starburst nucleus of the 16
sources with (O/H) $< 0.4$ (O/H)$_\odot$ and only 3 HII galaxies among
the 14 objects with (O/H) $> 0.6$ (O/H)$_\odot$. For Sample II, the
only two Starburst nuclei are also the most metal rich objects.
Furthermore, the four (O/H) = 0.4--0.6 (O/H)$_\odot$ sources (open
circles) located below the HII galaxy sequences in the $Z$-dependent
Figs.~\ref{fig:Excitation_Raimann}a and b are precisely the four
intermediate Starburst/HII galaxy groups defined by Raimann \etal
(2000a).  Metallicity and activity class are hence practically
equivalent quantities.

Figs.~\ref{fig:Excitation_SBMQ} and \ref{fig:Excitation_Raimann} show
the behavior of [OIII]/H$\beta$ and [OIII]/[OII] against
$\ov{t}_{SB}$ for Samples I and II respectively. In this section we
discuss only results for {\it metal poor} objects (triangles and
circles), which are mostly HII galaxies. These systems present clear
trends of decreasing excitation for increasing $\ov{t}_{SB}$, in
qualitative agreement with theoretical predictions. Sources in Sample
II join smoothly the sequences defined by sources in Sample I in all
plots above, extending it to smaller $\ov{t}_{SB}$. As already
explained, this happens mainly because of its objective prism
selection, which favors the detection of young starbursts (e.g.\
Stasi\'nska \& Leitherer 1996). Since emission lines are powered
solely by the most massive stars, they should be insensitive to the
presence of older, non-ionizing populations, and hence we can expect
the decrease in gas excitation to level off for $\ov{t}_{SB} > 10^7$
yr. This is consistent with the distributions of low Z objects in
Figs.~\ref{fig:Excitation_SBMQ} and \ref{fig:Excitation_Raimann}.  We
have also investigated other line ratios, such as [OII]/H$\beta$ and
[NII]/H$\alpha$, both of which increase systematically with increasing
$\ov{t}_{SB}$.

These same trends, were identified and discussed by Stasi\'nska \etal
(2001), who used $W_{H\beta}$ as an age indicator. This agreement is
hardly surprising, since we have empirically {\it verified} that
$W_{H\beta}$ and our evolutionary index $\ov{t}_{SB}$ are
related. In fact, as a corollary of this relation, we can automatically
subscribe all trends found using $W_{H\beta}$ as measure of evolution!
We therefore need not repeat here the extensive discussions on the
evolution of emission line properties of starbursts by Stasi\'nska
\etal (2001) and previous studies. The usual caveats about reddening
sensitive line ratios (such as [OIII]/[OII]) and the effects of an
absorption component in H$\beta$, discussed in the references above,
also apply here.

Of course, $\ov{t}_{SB}$ is a new age indicator, based entirely on
measured {\it stellar} properties. Sure enough, it too has its
limitations, but these are of a completely different nature than the
uncertainties affecting emission line age diagnostics (e.g., the
dilution correction for $W_{H\beta}$ or $W_{[OIII]}$). This reassuring
agreement supports the interpretation of the trends in [OIII]/H$\beta$
and [OIII]/[OII] against $\ov{t}_{SB}$ for metal poor objects as a
result of (theoretically expected) evolution of the gas excitation.

\subsection{Metallicity effects}

\label{sec:Metallicity}

%***FIG***FIG***FIG***FIG***FIG***FIG***FIG***FIG***FIG***FIG***FIG
\begin{figure}
\resizebox{9cm}{!}{\includegraphics{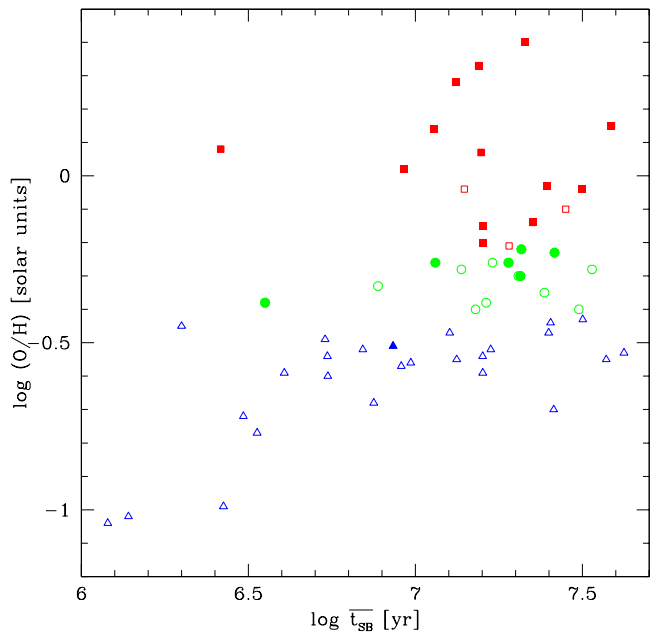}}
\caption{Gas metallicity against the evolutionary index
$\ov{t}_{SB}$. Symbols as in Fig.~\ref{fig:W_Hb_x_age_SBMQ}.
Starburst nuclei (filled symbols) are richer in metals than HII
galaxies, and their mixed populations of $10^6$--$10^8$ yr stars make
them look more evolved, producing the concentration towards to the top
right.}
\label{fig:Metallicity}
\end{figure}
%***FIG***FIG***FIG***FIG***FIG***FIG***FIG***FIG***FIG***FIG***FIG

Galaxies in our (O/H) $> 0.6$ (O/H)$_\odot$ metal-rich bin, are
heavily concentrated in the bottom-right regions of
Figs.~\ref{fig:W_Hb_x_age_SBMQ}--\ref{fig:Metallicity}, corresponding
to low emission line equivalent widths, low excitation and large age.
Despite their horizontal offset towards large $\ov{t}_{SB}$, high
$Z$ systems (plotted as squares) are well mixed with metal poor
systems in the $W_{H\beta}$ diagrams (Figs.~\ref{fig:W_Hb_x_age_SBMQ}
and \ref{fig:W_Hb_x_age_HIIgal}), in agreement with the idea that
$W_{H\beta}$ is largely insensitive to $Z$.  In the other diagrams,
however, high $Z$ galaxies are clearly offset along the vertical axis,
particularly in the gas-excitation plots
(Figs.~\ref{fig:Excitation_SBMQ} and
\ref{fig:Excitation_Raimann}). Furthermore, while metal poor objects
line up on broad but well defined sequences of decreasing
$W_{[OIII]}$, [OIII]/H$\beta$ and [OIII]/[OII] for increasing
$\ov{t}_{SB}$, no clear trends appear when considering high $Z$
objects by themselves.  The scattered distribution of metal rich
objects in these plots is in qualitative agreement with the models by
Stasi\'nska \etal (2001), which show that $Z$-dependent indices such
as those used in
Figs.~\ref{fig:W_Hb_x_age_SBMQ}--\ref{fig:Excitation_Raimann} are {\it
not} good chronometers for metal rich starbursts.

A more intriguing result is the systematic displacement of high $Z$
galaxies towards large $\ov{t}_{SB}$. A naive interpretation of this
offset would be that these systems represent the late stages of
evolution of metal poor starbursts. In this scenario, a triangle would
become a circle and then a square in
Figs.~\ref{fig:W_Hb_x_age_SBMQ}--\ref{fig:Excitation_Raimann}. Note,
however, that this would require that (O/H) increases by factors of
3--10 in less than $10^8$ yr. Furthermore, this scenario would not be
general, since there are several metal poor objects with large
$\ov{t}_{SB}$.  A more appropriate reading of
Figs.~\ref{fig:W_Hb_x_age_SBMQ}--\ref{fig:Excitation_Raimann} is that
there is a wide spread in $Z$ for evolved starbursts, but there is
practically no young metal rich system (the single exception being
group G\_Mrk710). This dichotomy is illustrated in
Fig.~\ref{fig:Metallicity}, where $Z$ is plotted against
$\ov{t}_{SB}$ for both samples.

We attribute this behavior to the fact that, as already explained,
metal rich sources are predominantly Starburst nuclei, whereas HII
galaxies dominate the lower $Z$ bins, as can be seen comparing the
location of filled and open symbols in Fig.~\ref{fig:Metallicity}. In
most HII galaxies the starburst population is dominated by the
youngest generations, partly due to selection effects (see Section
\ref{sec:W_Hb_corrections}) and partly due to the fact that these are
small galaxies, where a single burst can have a large impact. In fact,
instantaneous burst models often provide an acceptable description of
these systems, provided allowance is made for the presence of an old
underlying population (Mas-Hesse \& Kunth 1999). Starburst nuclei, on
the other hand, present a more even distribution of stellar ages in
the $10^6$--$10^8$ yr range (Lan\c{c}on \etal 2001), more compatible
with an extended star formation episode than with a coeval burst. This
age mixture is detected by the synthesis, resulting in a skew of the
$\ov{t}_{SB}$ index towards larger values.

We therefore conclude that the trend of $Z$ with $\ov{t}_{SB}$
simply reflects the fact that metal rich objects have a more complex
recent history of star formation than metal poor objects.

\section{Summary}

\label{sec:conclusions}

We have investigated the evolution of emission line properties in
Starburst nuclei and HII galaxies using age diagnostics based on their
observed integrated stellar population properties.  Our main results
can be divided in two parts.

In the first part of this paper, we have presented the results of an
empirical population synthesis (EPS) analysis of star-forming galaxies
and explored ways to condense these results onto simple diagrams and
indices designed to assess the evolutionary state of a stellar
population. Two useful tools were developed:

\begin{itemize}

\item[(1)] {\it An evolutionary diagram:} A compact description of
stellar populations in terms of young ($\le 10^7$ yr), intermediate
age ($10^8$ yr) and old ($\ge 10^9$ yr) components allows the
evolutionary state of a galaxy to be assessed by its location on a
$(x_Y,x_I,x_O)$ diagram, each axis carrying the contribution of stars
within a given age range to the total flux.

\item[(2)] {\it Mean ages:} Flux--weighted mean ages of both the total
stellar population ($\ov{t}$) and the starburst component
($\ov{t}_{SB}$) were defined .  

\end{itemize}

Both tools were tested with theoretical galaxy spectra for
instantaneous bursts and continuous star formation.  These tests
showed that the evolution of stellar populations is adequately mapped
by these empirical tools, supporting their application to real
galaxies. Perhaps the main conclusion here is that one can achieve a
good first order description of the evolutionary state of a starburst
using very little spectral information; our analysis used just 3
absorption lines plus 2 continuum colors in the 3600--4500 \AA\
interval.

The EPS-analysis of two samples of starbursting galaxies showed them
be distributed along the direction of evolution in the $(x_Y,x_I,x_O)$
diagram. This result encouraged us to use our mean starburst age
$\ov{t}_{SB}$ as an {\it empirical clock} to gauge the evolutionary
state of starbursts.

In the second part of this study we have investigated correlations
between the emission line properties of Starburst nuclei and HII
galaxies and the $\ov{t}_{SB}$ index in order to test, in a
completely empirical way, whether emission lines evolve along with the
stars in starbursts.  The results of this investigation can be
summarized as follows.

\begin{itemize}

\item[(1)] We have verified that the equivalent widths of H$\beta$ and
[OIII] decrease for increasing $\ov{t}_{SB}$. This is in accordance
with well known, but little tested, theoretical expectations.

\item[(2)] The use of $W_{H\beta}$ and $W_{[OIII]}$ as age indicators
is hampered by the the diluting effects of an old underlying stellar
population unrelated to the starburst. Besides providing a
quantitative assessment of evolution, the EPS analysis provides a
straight forward estimate of this effect. 

\item[(3)] As a whole, Starburst nuclei are found to have a more even
distribution of stellar ages in the $10^6$--$10^8$ yr range than HII
galaxies, which are often dominated by the youngest generations.

\item[(4)] Three Seyfert 2 objects were also analysed, two of which
have stellar population characteristics radically different from those
in starburst galaxies, as seen, for instance, by their location on the
evolutionary diagram. These two sources also have dilution-corrected
$W_{H\beta}$ values well above those of starbursts. The third object
has characteristics suggestive of a composite starburst + Seyfert 2
system.

\item[(5)] The gas excitation, as measured by emission line ratios,
was found to decrease systematically for increasing $\ov{t}_{SB}$,
also in agreement with theoretical predictions. This evolutionary
sequence is only well defined for metal poor objects, which are mostly
HII galaxies. Metal rich galaxies do not present clear evolutionary
trends in the gas excitation indices, in qualitative agreement with
photoionization models for evolving starbursts.

\end{itemize}

\section*{ACKNOWLEDGMENTS}

We thank Claus Leitherer, Daniel Raimann, Eduardo Telles and Henrique
Schmitt and for discussions and suggestions on an earlier version of
this manuscript. RRL and JRSL acknowledge post-graduate fellowships
awarded by CNPq. Support from CNPq, PRONEX are also acknowledged.

\end{document}